\documentclass[12pt]{iopart}

\usepackage{iopams}

\ifx\pdfoutput=0
  \usepackage[pdftex]{graphicx}
  \usepackage[pdftex,usenames]{color}
\else
  \usepackage[dvips]{graphicx}
  \usepackage[usenames]{color}
  \usepackage{epsfig}
  \epsfclipon
\fi

\newcommand{\diff}{{d}}
\newcommand{\ubar}{\bar u}
\newcommand{\dbar}{\bar d}
\newcommand{\qbar}{\bar q}
\newcommand{\fbar}{\bar f}
\newcommand{\xF}{x_F}

\begin{document}

\topical
[Exploring the Partonic Structure of Hadrons through the Drell-Yan Process]
{Exploring the Partonic Structure of Hadrons through the Drell-Yan Process}

\author{P.E.~Reimer}

\address{
Physics Division, Argonne National Laboratory, Argonne, IL, 60439, USA}
\ead{reimer@anl.gov}

\begin{abstract}
The Drell-Yan process is a standard tool for probing the partonic
structure of hadrons. Since the process proceeds through a
quark-antiquark annihilation, Drell-Yan scattering possesses a unique
ability to selectively probe sea distributions.  This review examines
the application of Drell-Yan scattering to elucidating the flavor
asymmetry of the nucleon's sea and nuclear modifications to the sea
quark distributions in unpolarized scattering.  Polarized beams and
targets add an exciting new dimension to Drell-Yan scattering.  In
particular, the two initial-state hadrons give Drell-Yan 
sensitivity to chirally-odd transversity distributions.
\end{abstract}


\submitto{\JPG}



\section{The Production of Massive Lepton Pairs}

\begin{figure}

  \center{\includegraphics{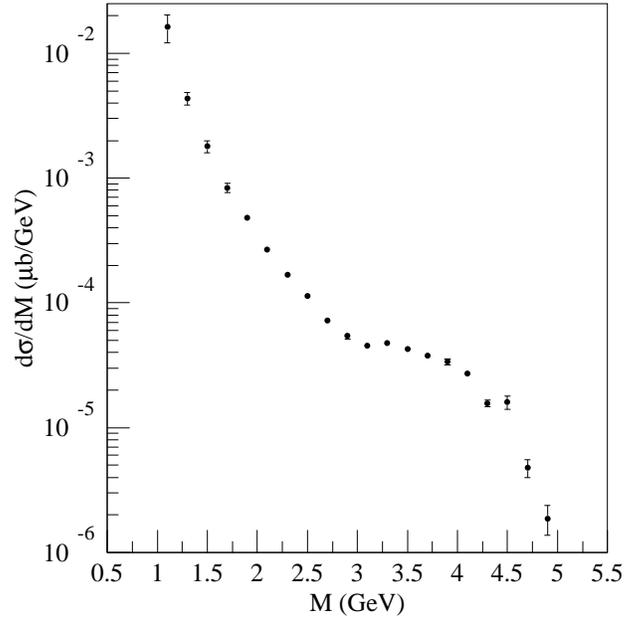}}

  \caption{The dilepton cross section as measured by Christenson {\it
  et al.}~\cite{PhysRevD.8.2016}, showing a rapid decrease as a
  function of dilepton mass.  The excess of events in the $3 <
  M_{\gamma^*} < 4$ region is from the dilepton decay of the
  $J/\psi$. \label{fig:christenson}}

\end{figure}

Sidney Drell and Tung-Mow Yan first proposed~\cite{PhysRevLett.25.316,
PhysRevLett.25.902.2} the process that now bears their names to
explain a continuum of massive lepton-antilepton pairs (dileptons)
that had been observed by Christenson {\it et al.} in proton-uranium
collisions at the Brookhaven AGS~\cite{PhysRevLett.25.1523,
PhysRevD.8.2016}.  The experiment was conducted to probe the large
momentum transfer region with time-like photons to complement
space-like measurements from lepton-proton deep inelastic scattering
(DIS) data and to search for new resonances.  A distinct feature of
these data was the rapid decrease in cross section as the mass of the
dilepton increased, as reproduced in figure~\ref{fig:christenson}.
While there were many plausible explanations of this spectra, it was
the mechanism proposed by S. Drell and T.-M. Yan which described the
spectra in terms of the (then very new) parton model of
Feynman~\cite{PhysRevLett.23.1415} that was eventually accepted.  In
this description, the dilepton cross section and its rapid decrease
with increasing dilepton mass was explained in terms of the
annihilation of a parton from one of the interacting hadrons with an
anti-parton from the other hadron.  The steeply falling cross section
was due the paucity of large-$x$ partons that are necessary to reach
high mass.

This article first reviews the basic formalism of the Drell-Yan
mechanism in unpolarized scattering and its relation to parton
distribution measurements.  Next, angular distributions of Drell-Yan
scattering and observed deviations from the expected distribution are
reviewed.  Finally longitudinally and transversely polarized Drell-Yan
measurements are discussed.  Within these discussions relevant recent
and proposed Drell-Yan measurements will be presented.

\section{The Drell-Yan Process}
\label{sec:unpolar}

\begin{figure}
  \center{\includegraphics{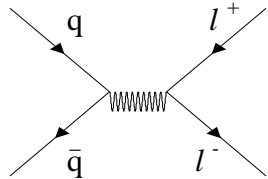}}

  \caption{Feynman diagram for the leading order Drell-Yan
  process. \label{fig:dylo}}
\end{figure}

\begin{figure}

  \center{\includegraphics{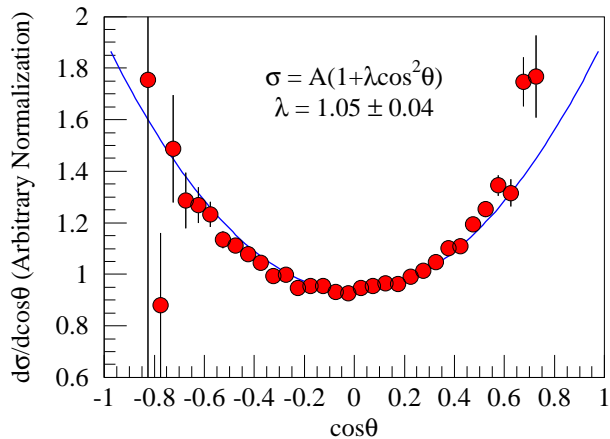}}

  \caption{The $\cos\theta$ dependence of the proton-proton Drell-Yan
  cross section as measured by the Fermilab E866/NuSea experiment.
  The curve shows the result of a fit of the data to $A\left( 1 +
  \lambda \cos^2 \theta
  \right)$~\cite{E866_angle}. \label{fig:costheta}}

\end{figure}

The production of massive dileptons through quark-antiquark
annihilation can be expressed in terms of a hard, short-distance
interaction term representing the cross section for quark-antiquark
annihilation into virtual photon and subsequent decay to a dilepton
pair, $\sigma_{q\bar q}$ (illustrated in figure~\ref{fig:dylo})
and the parton probability densities within the interacting hadrons.
The hard scattering cross section is given by
\begin{equation}
\sigma_{q\bar q} =\frac{4\pi \alpha^2}{3M_{\gamma^*}^2} \frac{1}{3} e_i^2,
\end{equation}
where the cross section is reduced by the final factor of $1/3$ since
the color-charge of the quark and antiquark must match, $e_i$ is the
fractional charge on the quark and $M_{\gamma^*}$ is the dilepton
mass.  To obtain the hadron-hadron cross section, it is necessary to
sum over the available quark flavors and account for the parton
distributions.  To leading order in the strong coupling constant,
$\alpha_s$, the Drell-Yan cross section is then
\begin{equation}
\fl \frac{\diff ^2\sigma}{\diff x_1\diff x_2} 
      = \frac{4\pi \alpha^2}{9M_{\gamma^*}^2} 
        \sum_{i} e_i^2
             \left[f_i(x_1,Q^2) \fbar_i(x_2,Q^2) 
                +  \fbar_i(x_1,Q^2) f_i(x_2,Q^2) \right],\label{eq:dylo}
\end{equation}
with the sum is over quark flavors, $i\in\{u, d, s, \ldots\}$.  The
parton distributions functions (PDFs) are given by $f_i(x,Q^2)$, where
$x$ is Bjorken-$x$ and $Q^2$ is the QCD scale at which the parton
distribution is probed.  In the case of Drell-Yan scattering, $Q^2 =
M^2_{\gamma^*}$.  (In general, $M^2_{\gamma^*}$ will be used when
discussing an invariant mass {\em measured} by an experiment and $Q^2$
will be used when discussing the QCD scale.)  The subscripts $1$ and
$2$ denote the interacting hadrons, which in a fixed target
experiment, are conventionally take as $1$ for the beam hadron and $2$ for the
target hadron.  Detailed derivations of this cross section may be found in
the literature~\cite{Kubar:1980zv, Kenyon:1982tg, Stirling:1993gc}.
The leading order Drell-Yan mechanism also predicts that the spin of
the virtual photon will be aligned providing a cross section that has
a $(1+\cos^2\theta)$ dependence, where $\theta$ is the polar angle
of the lepton in the rest frame of the virtual
photon~\cite{PhysRevLett.25.316}, in agreement with data as shown in
figure~\ref{fig:costheta}.  Additional features of the angular
distributions and their deviations from $(1+\cos^2\theta)$ are
discussed in section~\ref{sec:angle}.

Experimentally, one measures the momenta of the outgoing lepton and
antilepton, allowing for the reconstruction of the virtual photon's
mass, $M_{\gamma^*}$, longitudinal momentum, $p_l$ and transverse
momentum, $p_T$.  It is generally more convenient to use the
variables
\begin{equation}
  \tau = M_{\gamma^*}^2/s
\end{equation}
and
\begin{equation}
  y = \frac{1}{2}\ln\left(\frac{E+p_l}{E-p_l}\right)~\textrm{(rapidity)},
\end{equation}
where $s$ is the square of the center-of-mass energy of the
interacting hadrons and $E$ is the virtual photon's energy.  From
these, the momentum fractions $x_1$ and $x_2$ (Bjorken-$x$) of the
interacting partons are given by
\begin{equation}
  x_{1,2} = \left(\tau+\frac{p_T^2}{s}\right)^{1/2} e^{\pm y}
\end{equation}
and the difference (Feynman-$x$)
\begin{equation} 
  x_F \equiv \frac{2p_l}{\sqrt{s}} \approx x_1 - x_2
\end{equation}
In the limit of $p_T\rightarrow 0$ and large $\sqrt{s}$, this is
equivalent to $M_{\gamma^*}^2 = x_1 x_2 s$ and $x_F = 2p_l/\sqrt{s} =
x_1 - x_2$.  For a further discussion of the differences
see~\cite{Kubar:1980zv}.

\begin{figure}
  \center{\includegraphics{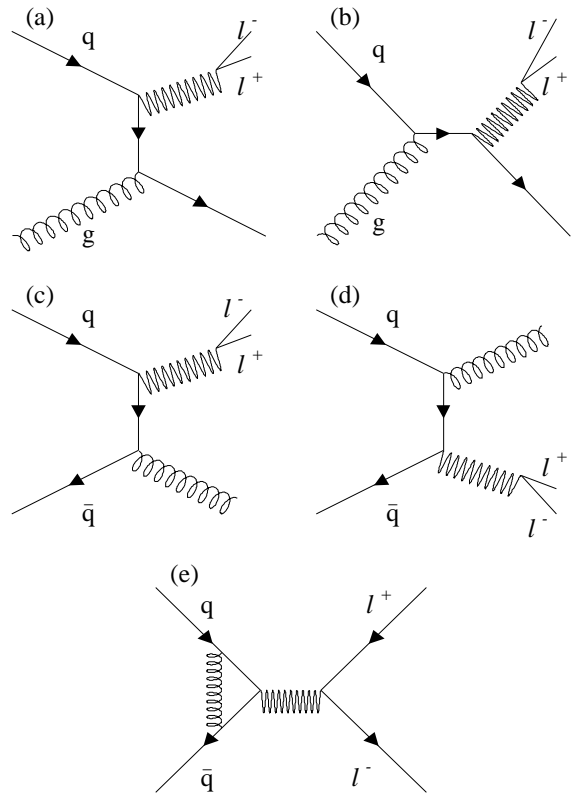}}

  \caption{Feynman diagram for the terms of next-to-leading order in
  $\alpha_s$ for the Drell-Yan process: (a) and (b) QCD Compton; (d)
  and (e) gluon production and (f) vertex
  correction~\cite{Kubar:1980zv}. \label{fig:allnlo}}
\end{figure}

\begin{figure}

  \center{\includegraphics[width=\columnwidth]{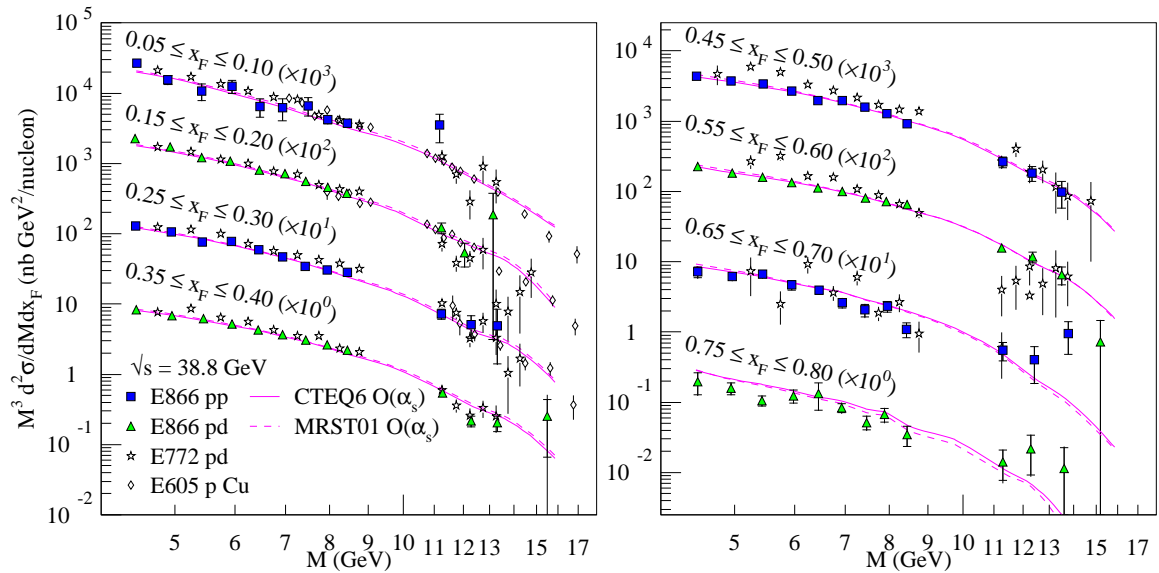}}

  \caption{Drell-Yan absolute cross sections measured by Fermilab
  E866/NuSea (proton-proton and proton deuterium)~\cite{Webb:2003bj,
  Webb:2003ps}, E772 (proton-deuterium)~\cite{PhysRevD.50.3038,
  PhysRevD.60.119903} and E605 (proton-copper)~\cite{PhysRevD.43.2815}
  compared with NLO cross section calculations based on the
  CTEQ6~\cite{Pumplin:2002vw} and MRST~\cite{Martin:2006qz} parton
  distributions.  There is an overall 6.5\% normalization uncertainty
  on the E866/NuSea data. \label{fig:absolute}}

\end{figure}

As experiments were able to reduce the systematic uncertainty of the
overall normalization of the Drell-Yan cross section, it became
apparent that, while the leading order cross section (using DIS PDFs)
explains many of the features of Drell-Yan scattering, it fails to
predict the overall magnitude of the cross section by a factor of
approximately 1.5-2.  This factor, traditionally known as the
``K''-factor, results from neglecting terms of higher order in
$\alpha_s$ in the cross section formula.  The next-to-leading order
(NLO) in $\alpha_s$ terms~\cite{Altarelli:1979ub, Harada:1979bj} of
the perturbative expansion, shown schematically in
figure~\ref{fig:allnlo}~\cite{Kubar:1980zv},
appear to account for the remainder of the
experimentally measured cross section, within the scale uncertainty of
the measurements and systematic uncertainty of the parton distribution
input to the calculations~\cite{Webb:2003bj, Webb:2003ps}, as shown in
figure~\ref{fig:absolute}.
(Recall, however, that while much of the global parton distribution
fits is dominated by DIS scattering results, Drell-Yan measurement
were also included in these fits.)  Higher order QCD corrections to
the cross section~\cite{PhysRevD.51.44, Hamberg:1990np}, as well as
electromagnetic radiative corrections to the cross section have also
been calculated~\cite{PhysRevD.57.199, PhysRevD.65.033007}.

The interpretation of the observed dilepton spectra in terms of parton
distributions relies on the factorization of the Drell-Yan cross
section into an infrared safe, short range hard scattering and the
parton distributions.  It further requires that these parton
distributions have the same meaning as DIS parton distributions. In a
twist expansion, the cross section can be expressed in terms of powers
of $1/(QR)$ where $Q^2$ is the hard scale and $R\approx\mathcal{O}
(1/\Lambda_\textrm{QCD})$ represents a non-perturbative
scale~\cite{Qiu:1990xx, Qiu:1990xy}.
\begin{equation}
\sigma_\textrm{DY} = \sigma_\textrm{Hard} + 
         \sum_n\mathcal{F}_{n=1}\left[1/(QR)^n\right],
\end{equation}
where $\sigma_\textrm{Hard}$ represents the convolution of the hard
scattering quark-antiquark cross section with the PDFs.  In leading
order in $\alpha_s$, $\sigma_\textrm{Hard}$ is given by
(\ref{eq:dylo}) but more generally includes higher powers of
$\alpha_s$. The sum over $\mathcal{F}_n$ represent terms of higher
twist.  Collins, Soper and Sterman~\cite{Collins:1985ue,
Collins:1988ig} and Bodwin~\cite{PhysRevD.31.2616} have shown that for
the leading twist term, $1/(QR)^0$, in this expansion the
contributions of non-factorisable soft gluons cancel and so the
leading twist term is factorisable.  Qiu and Sterman showed that
factorization can be extended to the $1/(QR)^2$ term for unpolarized
scattering and to $1/(QR)$ in a polarized asymmetry~\cite{Qiu:1990xx,
Qiu:1990xy}.  Factorization breaks down for terms in the expansion
beyond this, but is not generally a problem since the $Q^2$ of the
typical Drell-Yan experiment is greater than the $J/\psi$ mass
squared, suppressing higher powers of $1/(QR)$.  It may be possible,
however, to observe the effects of power corrections by investigating
regions in which the leading twist terms are suppressed such as
high-$x$.  Indeed, one possible explanation of an observed departure
at large-$x$ in pion-induced Drell-Yan from the $(1+\cos^2\theta)$
dependence~\cite{PhysRevLett.43.1219, Guanziroli:1987rp,
Falciano:1986wk, PhysRevD.39.92, PhysRevD.44.1909} is the presence of
higher twist terms as suggested by Berger and
Brodsky~\cite{PhysRevLett.42.940, Berger:1979xz}.  Drell-Yan angular
distributions are discussed in greater detail in
section~\ref{sec:angular}.  This explanation is consistent with the
work of Qiu and Sterman~\cite{Qiu:1990xy}.

In the $\tau \equiv M_{\gamma^*}^2/s \rightarrow 1$ limit, the cross
section is no longer adequately described by an expansion in
$\alpha_s$ to any fixed order.  In this regime, the energies of soft
and of collinear gluons are no longer negligible when compared with
the available energy in the system.  In the cross section, the
leading-logarithmic terms of the form
\begin{equation}
\alpha_s^k\frac{\ln^{m-1}\left(1-z\right)}{1-z}
    \textrm{~with~} \left(m \le 2k \right)
\end{equation}
are responsible for large corrections, where $z = \tau/(x_1 x_2)$
represents a partonic level version of $\tau$.  These terms must be
``resummed'' to all orders in $\alpha_s$ to adequately describe the
process.  Resummation was pioneered for the Drell-Yan process by
G. Sterman~\cite{Sterman:1986aj} and S. Gatani and
L. Trentadue~\cite{Catani:1989ne, Catani:1990rp}, where the
resummation is described in terms of a Mellin transformation.  A more
recent alternative approach by A. Idilbi and X. Ji for
Drell-Yan~\cite{Idilbi:2005ky, Idilbi:2006dg} uses soft-collinear
effective field theory, based on a similar description by
A. Manohar~\cite{Manohar:2003vb} for DIS, and arrives at the same
result.

The vast majority of Drell-Yan data is not near $\tau\rightarrow 1$
limit, which would require the interaction of very large-$x$ parton
from each hadron--a limit very difficult to reach when one of the
partons is a sea antiquark.  (Although, arguments have been made that
resummation should be considered in any case~\cite{Bolzoni:2006ky},
and the effects of neglecting resummation have been included estimates
of the uncertainties in parton distributions~\cite{Martin:2003sk}.)
Recently, the PAX collaboration~\cite{Barone:2005pu} has proposed
Drell-Yan measurements using an antiproton-proton collisions, thus
making available valence antiquarks at large-$x$.  In addition, the
center of mass energy range is relatively small, with $30 \lesssim s
\lesssim 200~\textrm{GeV}^2$.  In these kinematics, the effects
described by resummation will contribute substantially to the cross
section~\cite{Shimizu:2005fp}, but are well understood.

\section{Sea Quark Distributions from Unpolarized Drell-Yan Measurements}
\label{sec:sea}

In a fixed target environment, where the decay leptons are boosted far
forward, Drell-Yan scattering has a unique sensitivity to the
antiquark distribution of the target hadron. Combining this boost with
the acceptance of the typical dipole-based spectrometer restricts the
kinematic acceptance of the detector to $\xF \gtrsim 0$ and
consequently to very high values of $x_1$, where the sea quarks are
suppressed by several orders of magnitude compared with the valence
distributions.  These beam valence quarks must then annihilate with an
antiquark in the target, thus preferentially selecting the first term
in (\ref{eq:dylo}).  This feature has been used by several recent
experiments to study the sea quark distributions in the nucleon and in
nuclei.  Two new experiments have been proposed to extend these
measurements to larger values of $x_2$.

Both experiments are conceptually similar to earlier fixed target
Drell-Yan experiments.  The Fermilab E906
experiment~\cite{E906_proposal} will use a 120 GeV proton beam
extracted from the Fermilab Main Injector.  The experiment is
already approved and expects to begin data collection in 2009.  It
will have a kinematic coverage of $0.08 \lesssim x_2 \lesssim 0.45$.
At the JPARC facility, a similar experiment using a 50 GeV proton beam
has been proposed with kinematic coverage of $0.2 \lesssim x_2
\lesssim 0.6 $.  An initial program (JPARC Phase I) using a 30 GeV
beam to study $J/\psi$ physics was also
proposed~\cite{JPARC_Drell-Yan_proposal}.

\subsection{Isospin Symmetry of the Light Quark Sea}
\label{sec:dbub}

For many years, it was believed that the proton's sea quark
distributions were $\dbar$-$\ubar$ symmetric, arising from
approximately equal splitting of gluons into $d\dbar$ and $u\ubar$
pairs.  While there were indications of $\dbar\ne\ubar$ from Drell-Yan
data in the early 1980's~\cite{PhysRevD.23.604} it was observation of
a violation of the Gottfried Sum Rule~\cite{PhysRevLett.18.1174} in
muon DIS by the New Muon
Collaborations~\cite{PhysRevLett.66.2712,PhysRevD.50.R1} that forced
this belief to be reconsidered.  Because of its sensitivity to the
antiquark distributions Ellis and Stirling
suggested~\cite{Ellis:1990ti} using Drell-Yan as a probe of the light
quark flavor asymmetry with hydrogen and deuterium targets.  To
illustrate this sensitivity, in leading order with the $x_1\gg x_2$,
assuming charge symmetry and the dominance of the $u\ubar$
annihilation term, the ratio of the per nucleon proton-proton to
proton-deuterium Drell-Yan yields can be expressed as
\begin{equation}
\left.\frac{\sigma_{pd}}{2\sigma_{pp}}\right|_{x_1\gg x_2} = 
    \frac{1}{2}\left[1+\frac{\dbar(x_2)}{\ubar(x_2)} \right].
\label{eq:dbarubar}
\end{equation}
The NLO terms in the cross section provide a small correction to this
{\em ratio} and are considered in the analysis of the data, along with
the deviation from the $x_1\gg x_2$ limit.  Fermilab E772 used their
existing proton induced Drell-Yan data to compare the $W/C$ and
$W/^2H$ yields to extract upper limits on the isospin asymmetry for
$0.04\le x \le 0.27$~\cite{PhysRevLett.69.1726}.  The first dedicated
measurement of the $\dbar/\ubar$ asymmetry using Drell-Yan scattering
was made by the CERN NA51 experiment~\cite{Baldit:1994jk}.  The
acceptance of the NA51 toroid-based detector was such that the average
rapidity $\langle y \rangle = 0$ and $x_1 = x_2 = 0.18$. The asymmetry
measured by NA51 was
\begin{equation}
\fl A_\textrm{DY} = 2\frac{\sigma^{pp}}{\sigma^{pd}} - 1 = 
                   \frac{\sigma^{pp}-\sigma^{pn}}
                         {\sigma^{pp}+\sigma^{pn}}
              = -0.09\pm 0.02~\textrm{(stat.)}\pm 0.025~\textrm{(syst.)},
\end{equation}
with the second equality only valid if nuclear effects are ignored.
From this NA51 extracted
\begin{equation}
\left.\frac{\ubar}{\dbar}\right|_{x=0.18} = 
    0.51\pm 0.04~\textrm{(stat.)}\pm 0.05~\textrm{(syst.)},
\end{equation}
a clear signal for isospin symmetry violation in the sea antiquark
distributions.

\begin{figure}

  \center{\includegraphics[width=\textwidth]{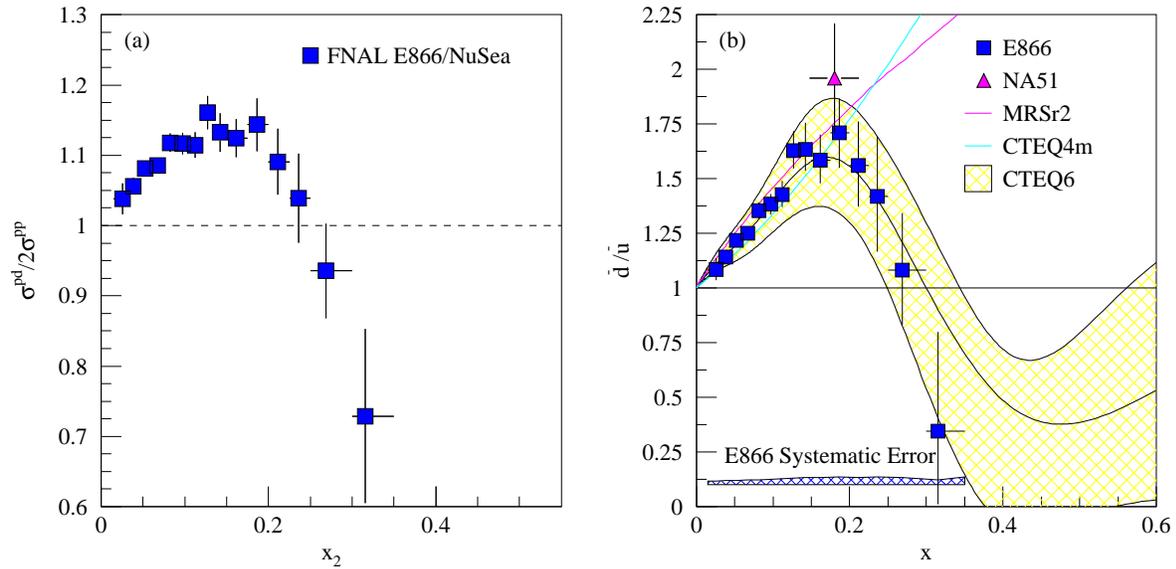}}

  \caption{(a) The blue squares show the ratio the proton-deuterium to
  twice the proton-proton Drell-Yan cross sections versus $x_2$ as
  measured by Fermilab E866/NuSea~\cite{Towell:2001nh}.  (b) The blue
  squares show $\dbar(x)/\ubar(x)$ ratio extracted by
  E866/NuSea~\cite{Towell:2001nh} The magenta triangle is the
  NA51~\cite{Baldit:1994jk} measurement of $\dbar/\ubar$.  The central
  curve in the cross filled band shows the $\dbar/\ubar$ ratio from
  the CTEQ6m fit~\cite{Pumplin:2002vw}, which included the E866/NuSea
  data, and the band represents the uncertainty from the fit.  The
  curves labeled CTEQ4M~\cite{Lai:1996mg} and
  MRS(r2)~\cite{Martin:1996as} show the parameterizations of
  $\dbar(x)/\ubar(x)$ which included the NA51 point and the EMC
  integral but not the E866/NuSea data. \label{fig:dbarubar}}

\end{figure}

The Fermilab E866/NuSea experiment used Drell-Yan to measure
$x$-dependence of the $\dbar/\ubar$ ratio. Fermilab E866/NuSea used a
spectrometer composed of three dipole magnets.  The first two magnets
served to focus large transverse momentum, $p_T$, dimuons into the
spectrometer while tracking surrounding the third magnet provided a
momentum measurement of the individual muons.  The experiment used 800
GeV protons extracted from the Fermilab Tevatron incident on hydrogen
and deuterium targets.  The remainder of the beam which did not
interact in the targets was intercepted by a copper beam dump
contained within the first magnet.  Additionally, the entire aperture
of the first dipole was filled with copper, carbon and borated
polyethylene, absorbing essentially all particles other than muons
produced in the interaction of the beam with the targets or beam
dump. (Fermilab E772~\cite{PhysRevLett.64.2479} used essentially the
same apparatus.)  E866/NuSea recorded 360,000 Drell-Yan events,
approximately two thirds from a deuterium target and the remainder
from a hydrogen target.  The ratio of Drell-Yan cross sections,
$\sigma^{pd}/\left(2\sigma^{pp}\right)$, measured by E866/NuSea as
well as the extracted ratio $\dbar(x)/\ubar(x)$ is shown in
figure~\ref{fig:dbarubar}.  When these cross section ratios were
included in global parton distribution fits~\cite{Pumplin:2002vw,
Martin:2004dh, Gluck:1998xa}, they completely changed the perception
of the sea quark distributions in the nucleon.

\begin{figure}

  \center{\includegraphics{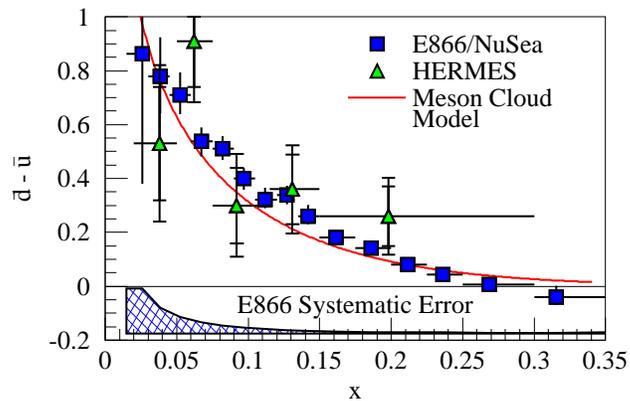}}

  \caption{The $\bar d(x) - \bar u(x)$ distribution as extracted by
  E866/NuSea (blue squares) using Drell-Yan~\cite{Towell:2001nh} and
  by HERMES (green triangles) using semi-inclusive
  DIS~\cite{PhysRevLett.81.5519}.  Also shown are a pion model
  calculation of Peng {\it et al.}~\cite{Peng:1998pa} based on the
  procedure of Kumano~\cite{PhysRevD.43.3067}.\label{fig:dmu}}

\end{figure}

The E866/NuSea data present an interesting picture of the sea quark
distributions of the nucleon that may shed some light on the origins
of the sea quarks.  At moderate values of $x$ the data show greater
than 60\% excess of $\dbar$ over $\ubar$, but as $x$ grows larger,
this excess disappears and the sea appears to be symmetric again.  If
the sea's origins are purely perturbative, then it is expected to have
only a {\em very} small asymmetry between $\dbar$ and
$\ubar$~\cite{Ross:1978xk, PhysRevC.55.900}. Many non-perturbative
explanations for the origin of the sea including meson cloud
models~\cite{Peng:1998pa, PhysRevD.43.3067}, chiral perturbation
theory~\cite{PhysRevD.45.2269, Szczurek:1993sc} or
instantons~\cite{Dorokhov:1991pv, Dorokhov:1993fc} have been suggested
which can explain a large asymmetry, but not the return to a symmetric
sea which is seen as $x\rightarrow 0.3$.  These models are reviewed
elsewhere~\cite{Garvey:2001yq}.  It is interesting to note the
importance of dynamical chiral symmetry breaking that Thomas {\it et
al.} have related to the $\dbar-\ubar$ difference through the presence
of Goldstone bosons which form a pion cloud around the
nucleon~\cite{PhysRevLett.85.2892}.  Henley {\it et al.} have
calculated the coordinate-space distribution of the $\dbar-\ubar$
asymmetry and observe that this distribution agrees with what is
expected if $\dbar-\ubar$ originates from a pion cloud surrounding the
nucleon~\cite{Henley:2000fm}.  Plotted in terms of $\dbar(x) -
\ubar(x)$, the observed E866/NuSea asymmetry, as shown in
figure~\ref{fig:dmu}~\cite{Towell:2001nh}, can be compared directly to
non-perturbative models since the (flavor symmetric) perturbative
component of the sea is removed.

The E866/NuSea data become less precise as $x$ increases beyond $0.25$
and the exact trend of $\dbar/\ubar$ is not clear.  To help understand
this region better, Fermilab E906 experiment has been approved to
collect Drell-Yan data in this region.  The Fermilab E906/Drell-Yan
experiment~\cite{E906_proposal} is modeled after its predecessors,
Fermilab E772 and E866/NuSea.  The E906 experiment will use a 120 GeV
proton beam rather than the 800 GeV beam used by E866 and E772.
Experimentally, the lower beam energy has two significant advantages.
First, the primary background in the experiment comes from $J/\psi$
decays, the cross section of which scales roughly with $s$.  The lower
beam energy implies less background rate in the spectrometer and
allows for a correspondingly higher instantaneous luminosity.  Second,
the Drell-Yan cross section at fixed $x_1$ and $x_2$ is inversely
proportional to $s$ (recall $M_{\gamma^*}^2\approx x_1x_2s$) and thus
the lower beam energy provides a larger cross section.  The muons
produced in a 120 GeV collision have a significantly smaller boost,
which forces the apparatus to be shortened considerably in order to
maintain the same $p_T$ acceptance.  The smaller boost will also
create a larger background from pion decay.  Fermilab E906 expects to
have statistical precision better than $1\%$ for $x < 0.35$ and $10\%$
for $0.35<x<0.45$, a clear improvement over the E866/NuSea data.

A similar experiment~\cite{JPARC_Drell-Yan_proposal} has also been
proposed for the JPARC facility.  This experiment would employ a 50
GeV proton beam with an apparatus similar in design to the
E906/Drell-Yan apparatus.  The initial phase of the JPARC facility is
for a 30 GeV synchrotron, which kinematicly has insufficient phase
space above the $J/\psi$ in mass for a Drell-Yan experiment; although,
once the entire facility is completed, including capabilities for 50
GeV beam, a significant Drell-Yan program could be mounted.

\subsection{Antiquark Distributions of Nuclei}
\label{sec:nuc_pi}

The distributions of partons within a free nucleon differ from those
of a nucleon bound within a heavy nucleus, an effect first discovered
by the European Muon Collaboration (EMC) in 1983~\cite{Aubert:1983xm}.
These nuclear effects are now generally divided into four regions in
$x$-space: the shadowing region with $x \lesssim 0.1$, the
anti-shadowing region covering $0.1\lesssim x \lesssim 0.3$, the EMC
effect region for $0.3\lesssim x\lesssim 0.6$ and finally a region
dominated by Fermi motion of the nucleons with $0.6 \lesssim x$. (For
a review of the nuclear EMC effect, see~\cite{Geesaman:1995yd}.)
Almost all of the data on nuclear dependencies is from charged lepton
DIS experiments, that are sensitive only to the charge-weighted sum of
all quark and antiquark distributions.  Nuclear effects in the sea
quark distributions may be entirely different from those in the
valence sector~\cite{Kulagin:2004ie}, but an electron or muon DIS
experiment would not be sensitive to this.  With the ability to probe
the antiquark distributions of the target, Drell-Yan presents an ideal
tool with which to distinguish between sea and valence effects.

\begin{figure}

  \center{\includegraphics[width=\columnwidth]{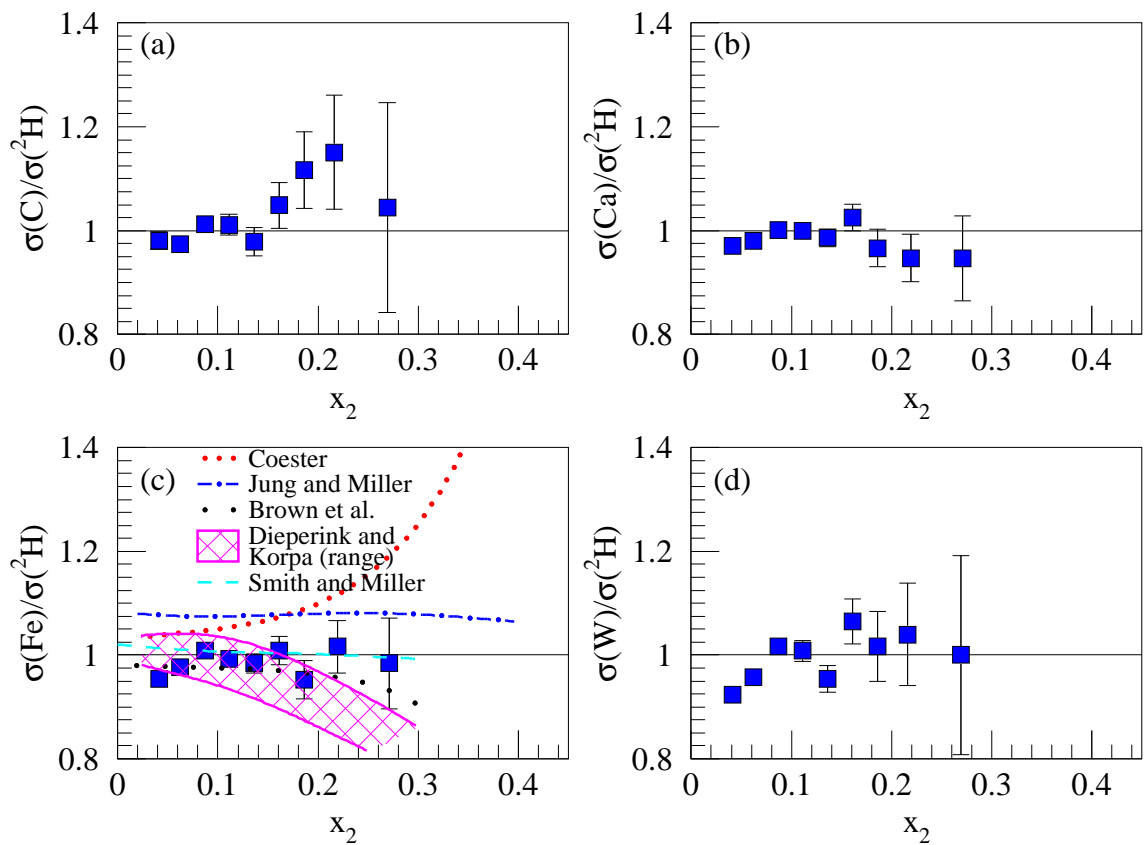}}

  \caption[]{E772 measurements of the ratio of Drell-Yan cross
  sections on (a) carbon, (b) calcium, (c) iron and (d) tungsten to
  deuterium \cite{PhysRevLett.64.2479}. Aside from shadowing in the
  $x_2 < 0.2$ region, no nuclear effects are observed. In (c), to
  illustrate the level of effects expected, curves based on several
  different representative models are also
  plotted~\cite{Berger:1983jk, PhysRevD.29.398, Coester_private,
  Jung:1990pu, Brown:1995dt, Dieperink:1997iv, Smith:2003hu}.  The
  differences between these curves is discussed briefly in the
  text. \label{fig:e772}}

\end{figure}

Some early models of the EMC effect were based on the convolution of a
virtual pion cloud with the nucleon.  In these models, virtual pion
contributions to nuclear structure functions were expected to lead to
sizable increases in sea distributions of the nuclei compared with
deuterium~\cite{Berger:1983jk, PhysRevD.29.398, Coester_private}.
Early Drell-Yan studies at both CERN~\cite{Bordalo:1987cs,
Bordalo:1987cr} and Fermilab~\cite{PhysRevD.23.604} lacked the
statistical sensitivity to observe the expected effects.  In 1990,
measurements by Fermilab E772~\cite{PhysRevLett.64.2479} found that
the expected enhancement was clearly {\em absent}, as shown in
figure~\ref{fig:e772}.  The non-observation of evidence of a pion
excess calls into question the most widely believed traditional
meson-exchange model~\cite{Carlson:1997qn} of the nucleus.  The
expected enhancement to the sea is illustrated in
Fig.~\ref{fig:e772}c, which shows the expected Drell-Yan ratio in iron
to deuterium, based on nuclear convolution model calculations by
Coester~\cite{Berger:1983jk, PhysRevD.29.398, Coester_private} meant
to explain the originally observed EMC effect.

More recent calculations~\cite{Jung:1990pu, Brown:1995dt,
Dieperink:1997iv, Smith:2003hu}, made in light of the E772 data,
predict a smaller nuclear dependence, consistent with the statistical
uncertainties of E772.  Jung and Miller~\cite{Jung:1990pu} revisited
the calculations of Berger and Coester and examined the effect of the
quantization of the pions on the light cone versus at ``equal time''.
With ``equal time'' quantization, they calculate a roughly flat 8\%
increase in the Drell-Yan iron cross section over the deuterium cross
section.  Brown {\it et al.}~\cite{Brown:1995dt} argue that with the
partial restoration of chiral symmetry, the masses of hadrons made up
of light quarks decrease with density.  This rescaling leads to
altered couplings, which lead to an overall $x$-dependent {\it
decrease} in the Drell-Yan cross section in nuclei.  Dieperink and
Korpa~\cite{Dieperink:1997iv} also argue that the Drell-Yan cross
section ratio should decrease in a nucleus.  Their arguments are based
on particle- and delta-hole model, which results in a strong
distortion of the free pion structure function.  Based on the chiral
quark soliton model, Smith and Miller predict essentially no nuclear
dependence for Drell-Yan, while at the same time qualitatively
explaining the observed EMC effect~\cite{Smith:2003hu}.  For $x>0.2$,
the E772 statistical uncertainties allow some freedom for these
models.  At $x\approx 0.3$ these newer models have nuclear effects of
the order 5 to 15\% in the Drell-Yan ratio and tend to diverge from
each other.  Fermilab E906 will provide the sensitivity needed to see
these differences with an expected statistical precision of 1\% at $x
= 0.3$ and data out to $x = 0.45$~\cite{E906_proposal}.

\section{Unpolarized Parton Distributions at Large-$x$ of the Beam Hadron}

In addition to probing the sea quark distribution of the target
nuclei, Drell-Yan scattering can be used to probe the structure of the
beam hadron.  As shown in (\ref{eq:dylo}), the cross section is a
convolution of the beam and target parton distributions.  The same
feature of the acceptance which allows fixed target experiments to
probe the target's sea quark distributions gives experimental access
to the high-$x$, valence parton distributions of the incoming hadron.
Absolute cross section measurements have been used to explore both
pionic and protonic parton distributions; although, some authors
suggest that there may be breakdown in factorization at
large-$x$~\cite{PhysRevLett.42.940, Berger:1979xz}.  Such effects
could cloud the partonic interpretation of these data.

The large-$x$ parton distributions of the proton are relatively
unknown, both in absolute magnitude and in the ratio of $d/u$.
Experimentally, these have been accessed through DIS.  Unfortunately,
much of the high-$x$ data used in DIS measurements involve nuclear
targets with relatively unknown and possibly large nuclear corrections
(even to deuterium)~\cite{Kuhlmann:1999sf}.  Proton induced Drell-Yan,
which is sensitive to the $4u+d$ distribution of the beam proton, is
an alternative way to reach high-$x$ with no nuclear corrections.  The
proton-proton and proton-deuterium cross section measurements of
Fermilab E866/NuSea when compared with NLO calculations based on both
the CTEQ6m~\cite{Pumplin:2002vw} and MRST~\cite{Martin:2004dh} parton
distributions show a small but systematic trend suggesting an
overestimation of the strength of this parton distribution at
large-$x$ ($x<0.8$ however)~\cite{Webb:2003bj}.  Future Drell-Yan
experiments such as Fermilab E906~\cite{E906_proposal} or a
JPARC-based experiment~\cite{JPARC_Drell-Yan_proposal} will be able to
improve dramatically on these measurements in both statistical
precision and reach in $x$.

The pion's parton distributions are of considerable interest because
of the pion's many unique roles in nuclear physics.  The pion's valence
antiquarks have been used to explain partially the observed
$\dbar/\ubar$ ratio in the proton through a pionic cloud around the
bare nucleon. (See Sec.~\ref{sec:dbub}.)  Models of nuclear binding
depend on pion exchange, and so the pionic quark distributions should
modify the parton distributions in nuclei. (See
Sec.~\ref{sec:nuc_pi}.)  While the pion is a $q\qbar$ system, its
extremely low mass arises from its role as the Goldstone boson of
dynamical chiral symmetry breaking and this role must be considered in
any descriptions of the pion's partonic nature.

There are a variety of theoretical and model-based descriptions of the
pion's valence parton structure.  At large-$x$ valence structure of
the pion can be parameterized as $(1-x)^\beta$, where QCD evolution of
the parton distributions makes $\beta$ a function of $Q^2$.  At low
$Q^2$, arguments based on the parton model~\cite{PhysRevLett.35.1416},
pQCD~\cite{Ji:2004hz, Brodsky:1995kg} and Dyson-Schwinger
equations~\cite{PhysRevC.63.025213} require that $\beta\approx 2$.  At
the same time the Drell-Yan-West relation~\cite{PhysRevLett.24.181,
PhysRevLett.24.1206}, duality arguments~\cite{Melnitchouk:2002gh} and
Nambu-Jona-Lasinio models~\cite{Shigetani:1993dx, Davidson:1995uv,
Weigel:1999pc, Bentz:1999gx} favor a linear dependence of
$\beta\approx 1$.  An early leading order analysis of pion induced
Drell-Yan data found $\beta = 1.26\pm 0.04$~\cite{PhysRevD.44.1909}
with $\langle M_{\gamma^*}\rangle = 5.2$~GeV.  Noting that the
strength of higher order diagrams could have a kinematic dependence, a
fit to the same data, this time in NLO found $\beta = 1.54\pm
0.08$~\cite{Wijesooriya:2005ir}, still somewhat in between a linear
and quadratic $1-x$ behavior.  Some future pionic Drell-Yan experiment
with improved kinematic resolution in the high-$x_1$ region could help
to resolve this question.

\section{Unpolarized Angular Distributions}
\label{sec:angle}
\label{sec:angular}

In leading order, ignoring transverse momenta ($k_T$) of the
interacting partons, the Drell-Yan angular distribution is naively
expected to have the form $\left(1+\cos^2\theta\right)$.  More
generally, Collins and Soper~\cite{PhysRevD.16.2219} have shown that
the expression for the angular distribution is
\begin{equation}
\frac{\diff\sigma}{\diff\Omega} \propto 
    1 + \lambda \cos^2\theta 
      + \mu\sin{2\theta}\cos{\phi} 
      + \frac{\nu}{2}\sin^2{\theta}\cos{2\phi},
\label{eq:angdist}
\end{equation}
where $\theta$ is the polar angle of the positive lepton in the rest frame
of the virtual photon and $\phi$ is the azimuthal angle.  The
additional terms arise from the $k_T$ of the interacting partons and
higher order graphs in $\alpha_s$.  After consideration of the
intrinsic $k_T$ of the partons, care must be taken in precisely
defining the the $z$-axis of rest frame of the virtual photon.  The
most common choices for this definition are the $u$-channel frame
where the the $z$-axis points anti-parallel to the target nucleon
direction; the Gottfried-Jackson frame
($t$-channel)~\cite{Gottfried:1964nx} has the $z$-axis pointing
parallel to the beam nucleon and the
Collins-Soper~\cite{PhysRevD.16.2219} frame where the $z$-axis bisects
the angle between the $u$-channel and Gottfried-Jackson $z$-axes, in
an attempt to minimize the effects of $k_T$ on the observed angular
distributions. The transformation between $\lambda$, $\mu$ and $\nu$
in the three frames is straight-forward~\cite{PhysRevD.39.92}.

In NLO, a relationship between $\lambda$ and $\nu$ of the
general angular distribution formula in (\ref{eq:angdist}) was
derived by C.S. Lam and W.-K. Tung~\cite{PhysRevD.21.2712}.  In
analogy to the Callan-Gross relationship of
DIS~\cite{PhysRevLett.22.156}, the Lam-Tung relation states that
\begin{equation}
1 - \lambda = 2 \nu. \label{eq:lam-tung}
\end{equation}
Unlike the Callan-Gross relation, the Lam-Tung relation is expected to
be largely unaffected by QCD~\cite{PhysRevD.21.2712}, including
resummation effects~\cite{Boer:2006eq}.

\begin{figure}

  \center{\includegraphics[width=\columnwidth]{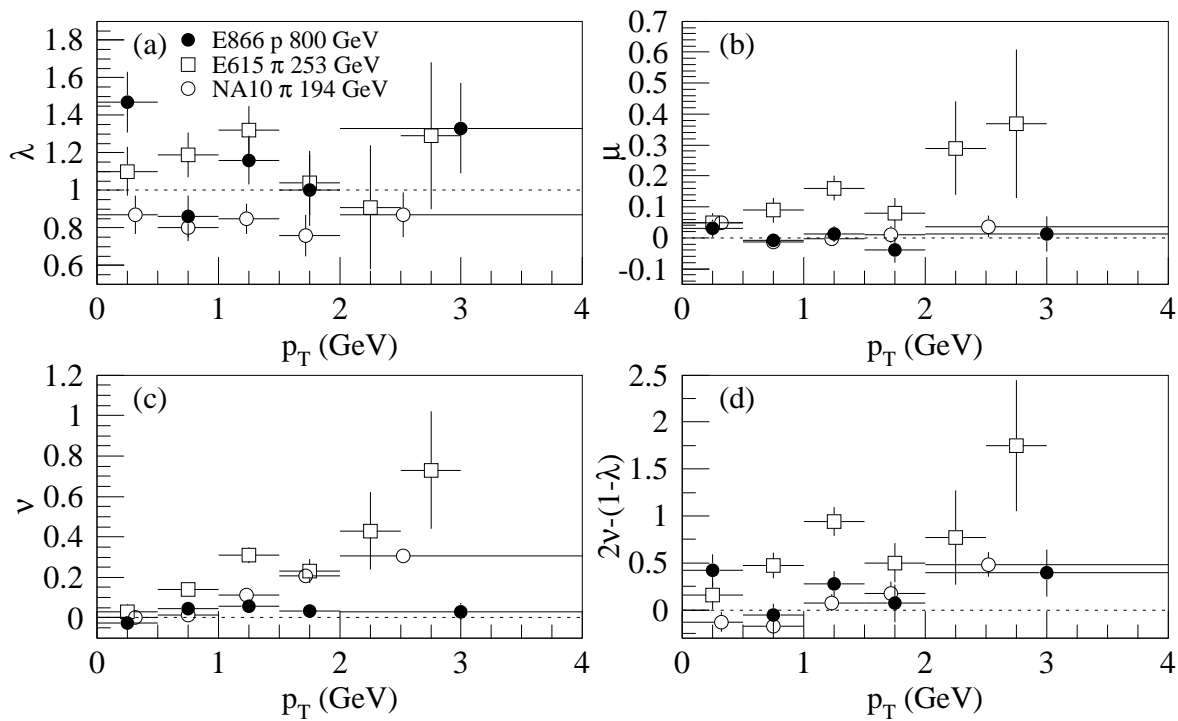}}

  \caption{The parameters (a) $\lambda$, (b) $\mu$ and (c) $\nu$ [see
  (\ref{eq:lam-tung})] of fits to the angular distributions as a
  function of $p_T$ from pion (CERN NA10~\cite{Falciano:1986wk,
  Guanziroli:1987rp} and Fermilab E615~\cite{PhysRevD.39.92,
  PhysRevD.44.1909}) and proton (Fermilab E866~\cite{Zhu:2006gx})
  induced Drell-Yan scattering. For the NA10 experiment, the data
  set with the best statistical precision, 194 GeV, is shown.  Plotted
  in (d) is $2\nu-(1-\lambda)$ which is predicted to be zero by the
  Lam-Tung relation.  For the pionic Drell-Yan data there is evidence
  for a violation of the Lam-Tung sum rule, especially at large-$p_T$.
  The uncertainties are statistical only and in (d) do not include
  possible correlations between $\lambda$ and $\nu$.\label{fig:angle}}

\end{figure}

The validity of the Lam-Tung relation has been studied with both
pion-induced Drell-Yan by CERN NA10~\cite{Falciano:1986wk,
Guanziroli:1987rp} and Fermilab E615~\cite{PhysRevD.39.92,
PhysRevD.44.1909} and proton induced Drell-Yan by Fermilab
E866/NuSea~\cite{Zhu:2006gx}.  Pionic Drell-Yan experiments have
observed a violation of the Lam-Tung relation.  This violation is most
prominent at high transverse momentum of the dilepton, $p_T$, where
$\nu$ rises without a corresponding decrease in $\lambda$, as shown in
figure~\ref{fig:angle}.  The violation appears to be independent of
the target nucleus~\cite{Guanziroli:1987rp}.  In contrast, the
recently released Fermilab E866 proton induced Drell-Yan data are
consistent with the Lam-Tung relation, even at
high-$p_T$~\cite{Zhu:2006gx}.

Berger and Brodsky have suggested that as $x\rightarrow 1$,
higher-twist effects cause the polarization of the virtual photon to
change from $\left(1+\cos^2\theta\right)$ to a $k_T$-dependent
$\sin^2\theta$~\cite{PhysRevLett.42.940}.  While effects consistent
with this have been observed in pionic Drell-Yan
data~\cite{PhysRevD.44.1909} at high-$x$, the violation of the
Lam-Tung relation is seen over a much broader range in $x$ than would
be expected by this explanation.  Brandenburg, Nachtmann and Mirke
hypothesized a spin correlation in the violation of factorization that
would give rise to a nonzero $\cos 2\phi$
distribution~\cite{Brandenburg:1993cj}, which could explain the pionic
Drell-Yan data.

An alternative explanation proposed by Boer~\cite{PhysRevD.60.014012}
based on the existence of a chiral-odd, $T$-odd distribution function,
$h_1^\perp(x, k_T)$, with an intrinsic transverse momentum dependence,
$k_T$, of Boer and Mulders~\cite{PhysRevD.57.5780}.  This {\em
distribution} function is an analog of the Collins {\em fragmentation}
function~\cite{Collins:1992kk}.  It represents the correlation of the
parton's transverse spin and $k_T$ in an unpolarized nucelon.  Boer
argues that the presence of the $h_1^\perp(x,k_T)$ distribution
function will induce a $\cos 2\phi$ dependence to the Drell-Yan cross
section and fits the observed NA10~\cite{Falciano:1986wk} data to a
crude model of this distribution function.  Within a quark
spectator-antiquark model, Lu and Ma have shown that the observed
$\cos 2\phi$ distribution can be reproduced with nonzero
$h_1^\perp(x,k_T)$ in both the pion and target
nucleon~\cite{Lu:2005rq}.  Other transversity distributions are
discussed again in section~\ref{sec:transversity}.

When considering any of these explanations for the violation of the
Lam-Tung relation in pion-induced Drell-Yan, it is important to
remember that these results must be reconciled with the apparent
absence of a violation in proton-induced Drell-Yan.  One significant
difference is that the valence antiquark in the pion-induced case
allows the experiment to probe the quark distributions of the target,
while in the proton induced case, only the target antiquark
distributions are probed.  Alternatively, the possible interpretation
as a higher-twist effect might have a kinematic dependence on
$\sqrt{s}$~\cite{Brandenburg:1993cj}.  Recall that the pionic data had
$\sqrt{s} = 11$ and $16$~GeV while the protonic data had $\sqrt{s} =
39$~GeV.  Such an effect should clearly be seen then in the upcoming
Fermilab E906 experiment with proton-induced Drell-Yan at $\sqrt{s} =
15$~GeV.

\section{Transversity Measurements with Drell-Yan Scattering}
\label{sec:transversity}

Polarized beams and targets add an extra dimension to Drell-Yan
scattering, enabling it to be used to study both quark helicity
(longitudinal) and transversity distributions. Close and Sivers
suggested that the sea quarks produced through gluon splitting would
be polarized and that this effect could be observed with
longitudinally polarized Drell-Yan
scattering~\cite{PhysRevLett.39.1116}.  Ralston and Soper considered
the possibility not only of longitudinally polarized but also of
transversely polarized scattering, proposing the measurement of
certain asymmetries as a test of the Drell-Yan
Model~\cite{Ralston:1979ys}.

In the near future, there are several proposed experiments at
polarized beam facilities.  One of the options that may be available
in Phase~II of the JPARC program would include a polarized proton
beam.  The FAIR program~\cite{fair} at GSI includes plans for a
polarized antiproton ring.  The PAX collaboration has
proposed~\cite{Barone:2005pu} a series of Drell-Yan and charm
measurements using this beam.  The first phase of the experiment would
involve polarized (or unpolarized) antiproton beams of up to 3.5 GeV
colliding with an internal polarized hydrogen target.  In the second
phase, the PAX experiment will be run as a collider experiment with up
to 15 GeV polarized antiprotons colliding with a beam of 3.5 GeV
polarized protons.  The RHIC-spin program also offers opportunities
for measuring polarized Drell-Yan scattering either with the existing
detectors or a new detector dedicated to polarized Drell-Yan
physics~\cite{RHICII, Perdekamp}.

In leading twist, in addition to the unpolarized parton distributions,
$f$, and the helicity distributions, $\Delta f$, a complete
description of the proton requires knowledge of the transversely
polarized parton distributions, denoted $h_1$.  (Various notations for
transversity distributions are used in the literature.  This notation
corresponds to that of Jaffe, Ji and
Mulders~\cite{Mulders:1995dh,PhysRevLett.67.552}.  Alternatively,
$\Delta_Tf$ in the notation of Barone and
Ratcliffe~\cite{Barone:2003fy,Barone:2001sp} or $\delta q$ have been
used.)  The quantity $h_1$ represents the net transverse helicity of
the quarks in a transversely polarized hadron. This distribution is
just as fundamental as the unpolarized and helicity distributions;
although, much less studied.  While the existence of the transverse
parton distributions was recognized in the late
1970's~\cite{Ralston:1979ys}, they received little consideration in
discussions of proton structure until recently.  This is, perhaps,
largely because of the difficulty in measuring these distributions.
Because transversity is a chirally odd quantity, it cannot be probed
in inclusive DIS, the traditional tool for deducing parton
distributions.  In order to observe $h_1$ chirality must be flipped
twice~\cite{Jaffe:1991ra}.  This requires two hadrons in the
interaction, for example, one each in the initial and final state or
two in the initial state.  Thus, both semi-inclusive DIS and Drell-Yan
scattering offer excellent opportunities to study transversity
distributions.  In the spin-$\frac{1}{2}$ nucleon, there is no gluon
transversity distribution and so the QCD evolution of $h_1$ is quite
different from the QCD evolution of $\Delta f$.  The distribution
$h_1$ evolves as a flavor non-singlet quantity~\cite{Barone:2001sp}.

With the additional consideration of intrinsic transverse momentum,
$k_T$, in the proton, two other distributions emerge.  The
Boer-Mulders function, $h^\perp_1(x,k_T)$~\cite{PhysRevD.57.5780}, was
introduced in section~\ref{sec:angular} to explain the observed $\cos
2\phi$ distributions in Drell-Yan scattering.  The Sivers
distributions function, $f^\perp_{1T}(x,k_T)$, characterized a
correlation in a polarized nucleon of the unpolarized parton density
with $k_T$~\cite{PhysRevD.41.83, PhysRevD.43.261}, and can be observed
in single spin asymmetries as discussed in section~\ref{sec:ssa}.
While these distributions functions vanish when integrated over $k_T$,
their possible existence and $k_T$-dependence offers a window into
spin-orbit correlations in the nucleon.  Recent measurements of
semi-inclusive DIS have provided an initial insight into both
transversity distributions and fragmentation functions; although
disentangling the effects of fragmentation from the distribution
functions can be difficult.  With Drell-Yan scattering, the
fragmentation functions are not relevant and one has a clean probe of
only the distribution functions.  For excellent reviews of the physics
of transversity, see~\cite{Barone:2003fy, Barone:2001sp}.

\subsection{Single Transverse Spin Asymmetries and the Sivers Function}
\label{sec:ssa}

In hadron-hadron interactions with one of the two hadrons is
transversely polarized and the other is unpolarized, surprisingly large
asymmetries of the form
\begin{equation}
A_N = \frac{d\sigma^\uparrow - d\sigma^\downarrow}
           {d\sigma^\uparrow + d\sigma^\downarrow}, \label{eq:ssa}
\end{equation}
have been observed ({\it e.g} the Fermilab E704
experiment~\cite{PhysRevLett.77.2626} which measured $p^\uparrow
p\rightarrow\pi X$).  D. Sivers suggested (prior to the publications
of the E704 results) that these single transverse-spin asymmetries
(SSA) could be used to provide information on the $k_T$-dependence of
unpolarized partons in a transversely polarized nucleon and that these
SSA were non-vanishing when the transverse motion of the individual
partons is considered~\cite{PhysRevD.41.83, PhysRevD.43.261}.  He
applied these argument to the reaction $hp^\uparrow\rightarrow\pi X$,
where $h$ represents any unpolarized hadron, $p^\uparrow$ is a
transversely polarized target and $\pi$ represents any detected
spinless final state meson.  The asymmetry arising from unpolarized
quarks in a transversely polarized hadron is known as the ``Sivers
asymmetry'' with the associated ``naively'' T-odd
$f^\perp_{1T}(x,k_T)$ parton distribution.  Initially, it was {\em
incorrectly} believed that $A_N$ would vanish because of time-reversal
invariance~\cite{Collins:1992kk}.  In QCD at leading twist it is also
expected to be zero~\cite{PhysRevLett.41.1689}. It was later shown
that with correct consideration of Wilson lines that the ``Sivers
asymmetry'' can be non-vanishing at {\em leading twist}.  The Wilson
lines provide for the gauge invariance in the parton number densities
and appear as soft initial-state interactions in
Drell-Yan~\cite{Collins:2002kn, Brodsky:2002cx}.  Alternatively, with
collinear QCD factorization at twist-three, it was recognized that
there could be a non-zero SSA~\cite{Efremov:1984ip,
PhysRevLett.67.2264}.  Both mechanisms, the $k_T$-dependent parton
distribution approach and twist-three QCD approach, provide mechanisms
to explain the same physical observable--a large SSA.  Within the
context of Drell-Yan scattering, the $k_T$-dependent parton
distributions are best applied to the domain in which $p_T \ll
M_{\gamma^*}$ while the twist-three QCD approach is more applicable
for $\Lambda_{\textrm{\tiny{QCD}}} \ll p_T$.  Recent work by Ji {\it
et al.}  has shown that these two approaches are in fact related, and
that they give the same result in the overlap region,
$\Lambda_{\textrm{\tiny{QCD}}}\ll p_T\ll
M_{\gamma^*}$~\cite{Ji:2006br, Ji:2006vf}.

The single spin asymmetries observed in the Fermilab E704
experiment~\cite{PhysRevLett.77.2626} could either be attributed to
the $f^\perp_{1T}(x, k_T)$ distribution in the proton or to a
transversity dependent ``Collins'' fragmentation
function~\cite{Collins:1992kk}.  While these two mechanism have
similar signatures in the E704 experiment, they produce different
angular distributions in semi-inclusive DIS.  Recently, the first
experimental evidence for both the Sivers $f^\perp_{1T}(x,k_T)$
distribution function and the Collins $H^\perp_1$ fragmentation
function has been observed by the HERMES
experiment~\cite{Airapetian:2004tw}.  Both the Sivers distribution and
Collins fragmentation functions have been extracted with global
analyses~\cite{Anselmino:2005ea, Vogelsang:2005cs} based on
observations from HERMES~\cite{Airapetian:2004tw} and
COMPASS~\cite{Alexakhin:2005iw}.  These analyses and model
calculations~\cite{Collins:2005rq} predict asymmetries of
approximately 8\% in $p^\uparrow\bar p$ Drell-Yan in the kinematics of
the proposed PAX experiment~\cite{Barone:2005pu}, well within their
statistical precision.  For $p^\uparrow p$ Drell-Yan at RHIC, these
analyses predict a 1--10\% asymmetry on the kinematics accepted.  One
very interesting consequence is that the Sivers asymmetry should have
the opposite sign when measured in Drell-Yan
scattering~\cite{Collins:2002kn}, {\it i.e.}
\begin{equation}
\left.f^\perp_{1T}(x,k_T)\right|_{\textrm{\tiny{DIS}}} = 
      -\left.f^\perp_{1T}(x,k_T)\right|_{\textrm{\tiny{DY}}}. 
      \label{eq:sivers}
\end{equation}
The experimental verification of (\ref{eq:sivers}) is a key to
validating the current understanding of $k_T$ effects in transversity
distributions and is one of the goals of the next generation of
transversely polarized Drell-Yan experiments.

\subsection{Double Spin Asymmetries}
\label{sec:double}

Doubly, transversely polarized Drell-Yan scattering offers the
possibility of measuring $h_1(x)$ without any complications from
fragmentation.  The asymmetry is given by~\cite{Ralston:1979ys,
Barone:2001sp, Barone:2003fy}
\begin{eqnarray}
A^\textrm{\tiny{DY}}_{TT} & = & 
              \frac{d\sigma^{\uparrow\uparrow} - d\sigma^{\uparrow\downarrow}}
                   {d\sigma^{\uparrow\uparrow} + d\sigma^{\uparrow\downarrow}} 
              \nonumber \\
 & = & \alpha_{TT} 
       \frac{\sum_i e_i^2 \left[h_{1i}(x_1)\bar{h}_{1i}(x_2) +
                              \bar{h}_{1i}(x_1)h_{1i}(x_2) \right]}
          {\sum_i e_i^2 \left[f_i(x_1)\bar{f}_i(x_2) + 
                              \bar{f}_i(x_1)f_i(x_2)\right]}, \label{eq:att}
\end{eqnarray}
where
\begin{equation}
  \alpha_{TT} = 
     \frac{\sin^2\theta\cos\left(2\phi\right)}
          {1+\cos^2\theta}.
\end{equation}
Note that this represents a convolution of the {\em quark} $h_1(x_1)$
and {\em antiquark} $\bar h_1(x_2)$ transversity distributions.
Unlike unpolarized Drell-Yan scattering, which also measured a
quark-antiquark convolution, the assumption that the quark
distributions are known from previous data (primarily DIS) is no
longer valid and so a significant amount of data is required to
untangle this convolution.  Furthermore, because the gluons do not
directly couple to $h_1$, it is expected that the sea contributions to
the asymmetry may be small.

As an alternative to this, the PAX collaboration~\cite{Barone:2005pu}
has proposed the measurement of antiproton-proton Drell-Yan scattering
in a collider mode.  In this case the cross section is dominated by
valence antiquarks annihilating with valence quarks.  Assuming the
dominance of the $u$-quark terms, the asymmetry in (\ref{eq:att})
reduces to~\cite{Barone:2005pu}
\begin{equation}
A^\textrm{\tiny{DY}}_{TT}  =  \alpha_{TT} 
       \frac{h_{1u}(x_1)h_{1u}(x_2)}
          {u(x_1)u(x_2)},
\end{equation}
thus giving a direct experimental access to $\left| h_{1u} \right|$.
Statistically, this is a challenging measurement, as polarized
antiprotons are difficult to create.  By relaxing the usual mass
requirement for Drell-Yan to include events both above and below the
$J/\psi$, PAX hopes to be able to obtain statistical precision of
10\%.  The relatively low energy of PAX places some data in a region
in which resummation may be important in interpreting the overall
cross sections.  Fortunately, it appears that the asymmetry
$A^\textrm{\tiny{DY}}_{TT}$ is largely unaffected by
resummation~\cite{Shimizu:2005fp}.

\section{Measurement of Quark Helicity with Longitudinally Polarized Drell-Yan}
\label{sec:longitudinally}

The total spin of the proton can be decomposed into the spin and
orbital contributions from the quarks and gluons:
\begin{equation}
\langle s_z^N\rangle = \frac{1}{2} = 
        \frac{1}{2}\Delta\Sigma + L_q + \Delta G + L_g, \label{eq:protonspin}
\end{equation}
where $\Delta \Sigma$ and $\Delta G$ represent the contributions of
the quark and gluon helicity respectively and $L_q$ and $L_g$ are the
orbital angular momentum of the quarks and gluons.  Motivated by the
observation of the European Muon Collaboration (EMC) that only a very
small fraction of the total spin of the proton is carried by the
quarks~\cite{Ashman:1987hv, Ashman:1989ig}, an analysis of the world's
data by the SMC group finds the fraction of the proton's spin carried
by quarks at $\Delta\Sigma = 0.38\pm 0.06$~\cite{PhysRevD.58.112002}.
The quark spin, $\Delta\Sigma$, can further be decomposed into
contributions of the individual flavors of quarks and antiquarks.

The HERMES experiment at DESY has been extremely successful at
studying the flavor decomposition of $\Delta\Sigma$ using
semi-inclusive DIS~\cite{Airapetian:2004mi}.  The results of this type
of experiment depend on precise knowledge of both the polarized and
unpolarized fragmentation functions.  Some authors have argued that
the contributions of the sea quarks to the polarized asymmetries are
small, reducing the sensitivity of this type of
experiment~\cite{Dressler:2000xj,Dressler:1999zg}. In Drell-Yan
scattering, however, this is not the case~\cite{Dressler:2000xj,
Dressler:1999zv}, and knowledge of the fragmentation functions is not
necessary.

As with unpolarized Drell-Yan, polarized Drell-Yan scattering offers a
window into the sea quark distributions.  The asymmetry for
longitudinally polarized Drell-Yan, $A^\textrm{\tiny{DY}}_{LL}$ is given
by~\cite{PhysRevLett.39.1116}
\begin{eqnarray}
A^\textrm{\tiny{DY}}_{LL} & = &
  \frac{d\sigma^{++}-d\sigma^{+-}}
       {d\sigma^{++}-d\sigma^{+-}} \nonumber \\
   & = & 
  \frac{\sum_i e_i^2\left[\Delta f_i(x_1)\Delta\bar f_i(x_2) 
                        + \Delta \bar f_i(x_1)\Delta f_i(x_2) \right]}
       {\sum_i e_i^2\left[f_i(x_1)\bar f_i(x_2) + \bar f_i(x_1)f_i(x_2) \right]}.
\label{eq:alongitudinal}
\end{eqnarray}
Here, $d\sigma^{++}$ ($d\sigma^{+-}$) denotes the spin parallel
(anti-parallel) Drell-Yan cross sections.  Making the same ``fixed
target'' assumptions as in the unpolarized case,
(\ref{eq:alongitudinal}) shows that $A^\textrm{\tiny{DY}}_{LL}$ measures
$\Delta\ubar$ in the target proton, convoluted with the large-$x$
$\Delta u$ distribution in the proton.  In the limit of exact SU(6),
Close and Sivers~\cite{PhysRevLett.39.1116} have shown that
\begin{equation}
A^\textrm{\tiny{DY}}_{LL}(x_1, x_2)  = \frac{1}{3}\frac{g_A}{g_V}
              \frac{\Delta \bar f(x_2)}{\bar f(x_2)}.
\end{equation}
With the addition of a polarized deuterium target, it would be
possible to extract $\Delta\dbar$ as well, providing that the valence
spin distributions are known.  Such a measurement has already been
proposed for the J-PARC Phase II facility with a 50 GeV polarized
proton beam.  The proposed experiment expects to be able to achieve
10\% statistical precision for $0.3\le x\le
0.5$~\cite{JPARC_Drell-Yan_proposal}.

\section{Conclusions}
\label{sec:conclusions}

Over the last 35 years, Drell-Yan scattering has played an important
role in elucidating the hadronic structure and will continue to do so
with the arrival of several new polarized and unpolarized experiments.
Unpolarized Drell-Yan scattering has been critical in measuring the
sea quark structure of the nucleon and the nucleus.  Fermilab
E866/NuSea has probed the flavor asymmetry of the antiquark sea,
showing conclusively that non-perturbative processes contribute to the
sea.  Fermilab E772 measured the nuclear effects on the sea quark
distributions, finding that widely accepted models of nuclear binding
to be lacking.  With transversely polarized beams, these experiments
will enable the extraction of the $h_1(x)$ transversity structure
function. In combination with DIS scattering, transverse single spin
asymmetries will test the understanding of the $k_T$ dependent Sivers
function, $f^\perp_{1T}(x,k_T)$.

\ack

This review has benefited greatly from suggestions and careful review
by J.-C. Peng and J. Qiu.  This work was supported the U.S. Department
of Energy, Office of Nuclear Physics, under Contract
No. DE-AC-02-06CH11357.

\section*{References}
\bibliographystyle{iopart-num}
\bibliography{Drell-Yan}

\providecommand{\newblock}{}
\begin{thebibliography}{100}
\expandafter\ifx\csname url\endcsname\relax
  \def\url#1{{\tt #1}}\fi
\expandafter\ifx\csname urlprefix\endcsname\relax\def\urlprefix{URL }\fi
\providecommand{\eprint}[2][]{\url{#2}}

\bibitem{PhysRevLett.25.316}
Drell S~D and Yan T~M 1970 {\em Phys. Rev. Lett.\/} {\bf 25}(5) 316--320

\bibitem{PhysRevLett.25.902.2}
Drell S~D and Yan T~M 1970 {\em Phys. Rev. Lett.\/} {\bf 25}(13) 902

\bibitem{PhysRevLett.25.1523}
Christenson J~H, Hicks G~S, Lederman L~M, Limon P~J, Pope B~G and Zavattini E
  1970 {\em Phys. Rev. Lett.\/} {\bf 25}(21) 1523--1526

\bibitem{PhysRevD.8.2016}
Christenson J~H, Hicks G~S, Lederman L~M, Limon P~J, Pope B~G and Zavattini E
  1973 {\em Phys. Rev. D\/} {\bf 8}(7) 2016--2034

\bibitem{PhysRevLett.23.1415}
Feynman R~P 1969 {\em Phys. Rev. Lett.\/} {\bf 23}(24) 1415--1417

\bibitem{Kubar:1980zv}
Kubar J, Le~Bellac M, Meunier J~L and Plaut G 1980 {\em Nucl. Phys.\/} {\bf
  B175} 251

\bibitem{Kenyon:1982tg}
Kenyon I~R 1982 {\em Rept. Prog. Phys.\/} {\bf 45} 1261

\bibitem{Stirling:1993gc}
Stirling W~J and Whalley M~R 1993 {\em J. Phys.\/} {\bf G19} D1--D102

\bibitem{E866_angle}
Reimer P (Fermilab E866/NuSea) 2007  Unpublished

\bibitem{Altarelli:1979ub}
Altarelli G, Ellis R~K and Martinelli G 1979 {\em Nucl. Phys.\/} {\bf B157} 461

\bibitem{Harada:1979bj}
Harada K, Kaneko T and Sakai N 1979 {\em Nucl. Phys.\/} {\bf B155} 169

\bibitem{Webb:2003bj}
Webb J~C 2003 {\em Measurement of continuum dimuon production in 800-GeV/c
  proton nucleon collisions\/} Ph.D. thesis New Mexico State University
  (\textit{Preprint} \eprint{hep-ex/0301031})

\bibitem{Webb:2003ps}
Webb J~C {\em et~al.\/} (E866/NuSea) 2003  (\textit{Preprint}
  \eprint{hep-ex/0302019})

\bibitem{PhysRevD.50.3038}
McGaughey P~L, Moss J~M, Alde D~M, Baer H~W, Carey T~A, Garvey G~T, Klein A,
  Lee C, Leitch M~J, Lillberg J, Mishra C~S, Peng J~C, Brown C~N, Cooper W~E,
  Hsiung Y~B, Adams M~R, Guo R, Kaplan D~M, McCarthy R~L, Danner G, Wang M,
  Barlett M and Hoffmann G 1994 {\em Phys. Rev. D\/} {\bf 50}(5) 3038--3045

\bibitem{PhysRevD.60.119903}
McGaughey P~L, Moss J~M, Alde D~M, Baer H~W, Carey T~A, Garvey G~T, Klein A,
  Lee C, Leitch M~J, Lillberg J, Mishra C~S, Brown C~N, Cooper W~E, Hsiung Y~B,
  Adams M~R, Guo R, Kaplan D~M, McCarthy R~L, Danner G, Wang M, Barlett M and
  Hoffmann G 1999 {\em Phys. Rev. D\/} {\bf 60}(11) 119903

\bibitem{PhysRevD.43.2815}
Moreno G, Brown C~N, Cooper W~E, Finley D, Hsiung Y~B, Jonckheere A~M, Jostlein
  H, Kaplan D~M, Lederman L~M, Hemmi Y, Imai K, Miyake K, Nakamura T, Sasao N,
  Tamura N, Yoshida T, Maki A, Sakai Y, Gray R, Luk K~B, Rutherfoord J~P,
  Straub P~B, Williams R~W, Young K~K, Adams M~R, Glass H and Jaffe D 1991 {\em
  Phys. Rev. D\/} {\bf 43}(9) 2815--2835

\bibitem{Pumplin:2002vw}
Pumplin J {\em et~al.\/} 2002 {\em JHEP\/} {\bf 07} 012 (\textit{Preprint}
  \eprint{hep-ph/0201195})

\bibitem{Martin:2006qz}
Martin A~D, Stirling W~J and Thorne R~S 2006 {\em Phys. Lett.\/} {\bf B636}
  259--264 (\textit{Preprint} \eprint{hep-ph/0603143})

\bibitem{PhysRevD.51.44}
Rijken P~J and van Neerven W~L 1995 {\em Phys. Rev. D\/} {\bf 51}(1) 44--63

\bibitem{Hamberg:1990np}
Hamberg R, van Neerven W~L and Matsuura T 1991 {\em Nucl. Phys.\/} {\bf B359}
  343--405

\bibitem{PhysRevD.57.199}
Baur U, Keller S and Sakumoto W~K 1998 {\em Phys. Rev. D\/} {\bf 57}(1)
  199--215

\bibitem{PhysRevD.65.033007}
Baur U, Brein O, Hollik W, Schappacher C and Wackeroth D 2002 {\em Phys. Rev.
  D\/} {\bf 65}(3) 033007

\bibitem{Qiu:1990xx}
Qiu J~w and Sterman G 1991 {\em Nucl. Phys.\/} {\bf B353} 105--136

\bibitem{Qiu:1990xy}
Qiu J~w and Sterman G 1991 {\em Nucl. Phys.\/} {\bf B353} 137--164

\bibitem{Collins:1985ue}
Collins J~C, Soper D~E and Sterman G 1985 {\em Nucl. Phys.\/} {\bf B261} 104

\bibitem{Collins:1988ig}
Collins J~C, Soper D~E and Sterman G 1988 {\em Nucl. Phys.\/} {\bf B308} 833

\bibitem{PhysRevD.31.2616}
Bodwin G~T 1985 {\em Phys. Rev. D\/} {\bf 31}(10) 2616--2642

\bibitem{PhysRevLett.43.1219}
Anderson K~J, Coleman R~N, Karhi K~P, Newman C~B, Pilcher J~E, Rosenberg E~I,
  Thaler J~J, Hogan G~E, McDonald K~T, Sanders G~H and Smith A~J~S 1979 {\em
  Phys. Rev. Lett.\/} {\bf 43}(17) 1219--1222

\bibitem{Guanziroli:1987rp}
Guanziroli M {\em et~al.\/} (NA10) 1988 {\em Z. Phys.\/} {\bf C37} 545

\bibitem{Falciano:1986wk}
Falciano S {\em et~al.\/} (NA10) 1986 {\em Z. Phys.\/} {\bf C31} 513

\bibitem{PhysRevD.39.92}
Conway J~S, Adolphsen C~E, Alexander J~P, Anderson K~J, Heinrich J~G, Pilcher
  J~E, Possoz A, Rosenberg E~I, Biino C, Greenhalgh J~F, Louis W~C, McDonald
  K~T, Palestini S, Shoemaker F~C and Smith A~J~S (Fermilab E615) 1989 {\em
  Phys. Rev. D\/} {\bf 39}(1) 92

\bibitem{PhysRevD.44.1909}
Heinrich J~G, Biino C, Greenhalgh J~F, Louis W~C, McDonald K~T, Palestini S,
  Russell D~P, Shoemaker F~C, Smith A~J~S, Adolphsen C~E, Alexander J~P,
  Anderson K~J, Conway J~S, Pilcher J~E, Possoz A and Rosenberg E~I (Fermilab
  E615) 1991 {\em Phys. Rev. D\/} {\bf 44}(7) 1909--1932

\bibitem{PhysRevLett.42.940}
Berger E~L and Brodsky S~J 1979 {\em Phys. Rev. Lett.\/} {\bf 42}(15) 940--944

\bibitem{Berger:1979xz}
Berger E~L 1980 {\em Z. Phys.\/} {\bf C4} 289

\bibitem{Sterman:1986aj}
Sterman G 1987 {\em Nucl. Phys.\/} {\bf B281} 310

\bibitem{Catani:1989ne}
Catani S and Trentadue L 1989 {\em Nucl. Phys.\/} {\bf B327} 323

\bibitem{Catani:1990rp}
Catani S and Trentadue L 1991 {\em Nucl. Phys.\/} {\bf B353} 183--186

\bibitem{Idilbi:2005ky}
Idilbi A and Ji X~d 2005 {\em Phys. Rev.\/} {\bf D72} 054016 (\textit{Preprint}
  \eprint{hep-ph/0501006})

\bibitem{Idilbi:2006dg}
Idilbi A, Ji X~d and Yuan F 2006 {\em Nucl. Phys.\/} {\bf B753} 42--68
  (\textit{Preprint} \eprint{hep-ph/0605068})

\bibitem{Manohar:2003vb}
Manohar A~V 2003 {\em Phys. Rev.\/} {\bf D68} 114019 (\textit{Preprint}
  \eprint{hep-ph/0309176})

\bibitem{Bolzoni:2006ky}
Bolzoni P 2006 {\em Phys. Lett.\/} {\bf B643} 325--330 (\textit{Preprint}
  \eprint{hep-ph/0609073})

\bibitem{Martin:2003sk}
Martin A~D, Roberts R~G, Stirling W~J and Thorne R~S 2004 {\em Eur. Phys. J.\/}
  {\bf C35} 325--348 (\textit{Preprint} \eprint{hep-ph/0308087})

\bibitem{Barone:2005pu}
Barone V {\em et~al.\/} (PAX) 2005  (\textit{Preprint} \eprint{hep-ex/0505054})

\bibitem{Shimizu:2005fp}
Shimizu H, Sterman G, Vogelsang W and Yokoya H 2005 {\em Phys. Rev.\/} {\bf
  D71} 114007 (\textit{Preprint} \eprint{hep-ph/0503270})

\bibitem{E906_proposal}
Reimer P, Geesaman D {\em et~al.\/} (Fermilab E906/Drell-Yan) 2001  Fermilab
  Proposal 906, Unpublished

\bibitem{JPARC_Drell-Yan_proposal}
Peng J~C, Sawada S {\em et~al.\/} 2006  JPARC Proposal, Unpublished

\bibitem{PhysRevD.23.604}
Ito A~S, Fisk R~J, J\"ostlein H, Kaplan D~M, Herb S~W, Hom D~C, Lederman L~M,
  Snyder H~D, Yoh J~K, Brown B~C, Brown C~N, Innes W~R, Kephart R~D, Ueno K and
  Yamanouchi T 1981 {\em Phys. Rev. D\/} {\bf 23}(3) 604--633

\bibitem{PhysRevLett.18.1174}
Gottfried K 1967 {\em Phys. Rev. Lett.\/} {\bf 18}(25) 1174--1177

\bibitem{PhysRevLett.66.2712}
Amaudruz P, Arneodo M, Arvidson A, Badelek B, Baum G, Beaufays J, Bird I~G,
  Botje M, Broggini C, Br\"uckner W, Br\"ull A, Burger W~J, Ciborowski J,
  Crittenden R, van Dantzig R, D\"obbeling H, Domingo J, Drinkard J, Dzierba A,
  Engelien H, Ferrero M~I, Fluri M~L, Grafstrom P, von Harrach D, van~der
  Heijden M, Heusch C and Ingram Q 1991 {\em Phys. Rev. Lett.\/} {\bf 66}(21)
  2712--2715

\bibitem{PhysRevD.50.R1}
Arneodo M, Arvidson A, Badelek B, Ballintijn M, Baum G, Beaufays J, Bird I~G,
  Bj\"orkholm P, Botje M, Broggini C, Br\"uckner W, Br\"ull A, Burger W~J,
  Ciborowski J, van Dantzig R, Dyring A, Engelien H, Ferrero M~I, Fluri L, Gaul
  U, Granier T, von Harrach D, van~der Heijden M, Heusch C, Ingram Q,
  Janson-Prytz K and de~Jong M 1994 {\em Phys. Rev. D\/} {\bf 50}(1) R1--R3

\bibitem{Ellis:1990ti}
Ellis S~D and Stirling W~J 1991 {\em Phys. Lett.\/} {\bf B256} 258--264

\bibitem{PhysRevLett.69.1726}
McGaughey P~L, Moss J~M, Alde D~M, Baer H~W, Carey T~A, Garvey G~T, Klein A,
  Lee C, Leitch M~J, Lillberg J~W, Mishra C~S, Peng J~C, Brown C~N, Cooper W~E,
  Hsiung Y~B, Adams M~R, Guo R, Kaplan D~M, McCarthy R~L, Danner G, Wang M~J,
  Barlett M~L and Hoffmann G~W 1992 {\em Phys. Rev. Lett.\/} {\bf 69}(12)
  1726--1728

\bibitem{Baldit:1994jk}
Baldit A {\em et~al.\/} (NA51) 1994 {\em Phys. Lett.\/} {\bf B332} 244--250

\bibitem{PhysRevLett.64.2479}
Alde D~M, Baer H~W, Carey T~A, Garvey G~T, Klein A, Lee C, Leitch M~J, Lillberg
  J~W, McGaughey P~L, Mishra C~S, Moss J~M, Peng J~C, Brown C~N, Cooper W~E,
  Hsiung Y~B, Adams M~R, Guo R, Kaplan D~M, McCarthy R~L, Danner G, Wang M~J,
  Barlett M~L and Hoffmann G~W 1990 {\em Phys. Rev. Lett.\/} {\bf 64}(21)
  2479--2482

\bibitem{Martin:2004dh}
Martin A~D, Roberts R~G, Stirling W~J and Thorne R~S 2005 {\em Eur. Phys. J.\/}
  {\bf C39} 155--161 (\textit{Preprint} \eprint{hep-ph/0411040})

\bibitem{Gluck:1998xa}
Gluck M, Reya E and Vogt A 1998 {\em Eur. Phys. J.\/} {\bf C5} 461--470
  (\textit{Preprint} \eprint{hep-ph/9806404})

\bibitem{Towell:2001nh}
Towell R~S {\em et~al.\/} (FNAL E866/NuSea) 2001 {\em Phys. Rev.\/} {\bf D64}
  052002 (\textit{Preprint} \eprint{hep-ex/0103030})

\bibitem{Lai:1996mg}
Lai H~L {\em et~al.\/} 1997 {\em Phys. Rev.\/} {\bf D55} 1280--1296
  (\textit{Preprint} \eprint{hep-ph/9606399})

\bibitem{Martin:1996as}
Martin A~D, Roberts R~G and Stirling W~J 1996 {\em Phys. Lett.\/} {\bf B387}
  419--426 (\textit{Preprint} \eprint{hep-ph/9606345})

\bibitem{Ross:1978xk}
Ross D~A and Sachrajda C~T 1979 {\em Nucl. Phys.\/} {\bf B149} 497

\bibitem{PhysRevC.55.900}
Steffens F~M and Thomas A~W 1997 {\em Phys. Rev. C\/} {\bf 55}(2) 900--908

\bibitem{Peng:1998pa}
Peng J~C {\em et~al.\/} (E866/NuSea) 1998 {\em Phys. Rev.\/} {\bf D58} 092004
  (\textit{Preprint} \eprint{hep-ph/9804288})

\bibitem{PhysRevD.43.3067}
Kumano S 1991 {\em Phys. Rev. D\/} {\bf 43}(9) 3067--3070

\bibitem{PhysRevD.45.2269}
Eichten E~J, Hinchliffe I and Quigg C 1992 {\em Phys. Rev. D\/} {\bf 45}(7)
  2269--2275

\bibitem{Szczurek:1993sc}
Szczurek A, Speth J and Garvey G~T 1994 {\em Nucl. Phys.\/} {\bf A570} 765--781

\bibitem{Dorokhov:1991pv}
Dorokhov A~E and Kochelev N~I 1991 {\em Phys. Lett.\/} {\bf B259} 335--339

\bibitem{Dorokhov:1993fc}
Dorokhov A~E and Kochelev N~I 1993 {\em Phys. Lett.\/} {\bf B304} 167--175

\bibitem{Garvey:2001yq}
Garvey G~T and Peng J~C 2001 {\em Prog. Part. Nucl. Phys.\/} {\bf 47} 203--243
  (\textit{Preprint} \eprint{nucl-ex/0109010})

\bibitem{PhysRevLett.85.2892}
Thomas A~W, Melnitchouk W and Steffens F~M 2000 {\em Phys. Rev. Lett.\/} {\bf
  85}(14) 2892--2894

\bibitem{Henley:2000fm}
Henley E~M, Renk T and Weise W 2001 {\em Phys. Lett.\/} {\bf B502} 99--103
  (\textit{Preprint} \eprint{hep-ph/0012045})

\bibitem{PhysRevLett.81.5519}
Ackerstaff K, Airapetian A, Akopov N, Akushevich I, Amarian M, Ashenauer E~C,
  Avakian H, Avakian R, Avetissian A, Bains B, Baumgarten C, Beckmann M,
  Belostotski S, Belz J~E, Benisch T, Bernreuther S, Bianchi N, Blouw J,
  B\"ottcher H, Borissov A, Brack J, Brauksiepe S, Braun B, Bray B, Brons S,
  Br\"uckner W and Br\"ull A 1998 {\em Phys. Rev. Lett.\/} {\bf 81}(25)
  5519--5523

\bibitem{Aubert:1983xm}
Aubert J~J {\em et~al.\/} (European Muon) 1983 {\em Phys. Lett.\/} {\bf B123}
  275

\bibitem{Geesaman:1995yd}
Geesaman D~F, Saito K and Thomas A~W 1995 {\em Ann. Rev. Nucl. Part. Sci.\/}
  {\bf 45} 337--390

\bibitem{Kulagin:2004ie}
Kulagin S~A and Petti R 2006 {\em Nucl. Phys.\/} {\bf A765} 126--187
  (\textit{Preprint} \eprint{hep-ph/0412425})

\bibitem{Berger:1983jk}
Berger E~L, Coester F and Wiringa R~B 1984 {\em Phys. Rev.\/} {\bf D29} 398

\bibitem{PhysRevD.29.398}
Berger E~L, Coester F and Wiringa R~B 1984 {\em Phys. Rev. D\/} {\bf 29}(3)
  398--411

\bibitem{Coester_private}
Coester F 2001 private communication

\bibitem{Bordalo:1987cs}
Bordalo P {\em et~al.\/} (NA10) 1987 {\em Phys. Lett.\/} {\bf B193} 368

\bibitem{Bordalo:1987cr}
Bordalo P {\em et~al.\/} (NA10) 1987 {\em Phys. Lett.\/} {\bf B193} 373

\bibitem{Carlson:1997qn}
Carlson J and Schiavilla R 1998 {\em Rev. Mod. Phys.\/} {\bf 70} 743--842

\bibitem{Jung:1990pu}
Jung H and Miller G~A 1990 {\em Phys. Rev.\/} {\bf C41} 659--664

\bibitem{Brown:1995dt}
Brown G~E and Rho M 1995 {\em Nucl. Phys.\/} {\bf A590} 527c--530c

\bibitem{Dieperink:1997iv}
Dieperink A~E~L and Korpa C~L 1997 {\em Phys. Rev.\/} {\bf C55} 2665--2674
  (\textit{Preprint} \eprint{nucl-th/9703025})

\bibitem{Smith:2003hu}
Smith J~R and Miller G~A 2003 {\em Phys. Rev. Lett.\/} {\bf 91} 212301
  (\textit{Preprint} \eprint{nucl-th/0308048})

\bibitem{Kuhlmann:1999sf}
Kuhlmann S {\em et~al.\/} 2000 {\em Phys. Lett.\/} {\bf B476} 291--296
  (\textit{Preprint} \eprint{hep-ph/9912283})

\bibitem{PhysRevLett.35.1416}
Farrar G~R and Jackson D~R 1975 {\em Phys. Rev. Lett.\/} {\bf 35}(21)
  1416--1419

\bibitem{Ji:2004hz}
Ji X~d, Ma J~P and Yuan F 2005 {\em Phys. Lett.\/} {\bf B610} 247--252
  (\textit{Preprint} \eprint{hep-ph/0411382})

\bibitem{Brodsky:1995kg}
Brodsky S~J, Burkardt M and Schmidt I 1995 {\em Nucl. Phys.\/} {\bf B441}
  197--214 (\textit{Preprint} \eprint{hep-ph/9401328})

\bibitem{PhysRevC.63.025213}
Hecht M~B, Roberts C~D and Schmidt S~M 2001 {\em Phys. Rev. C\/} {\bf 63}(2)
  025213

\bibitem{PhysRevLett.24.181}
Drell S~D and Yan T~M 1970 {\em Phys. Rev. Lett.\/} {\bf 24}(4) 181--186

\bibitem{PhysRevLett.24.1206}
West G~B 1970 {\em Phys. Rev. Lett.\/} {\bf 24}(21) 1206--1209

\bibitem{Melnitchouk:2002gh}
Melnitchouk W 2003 {\em Eur. Phys. J.\/} {\bf A17} 223--234 (\textit{Preprint}
  \eprint{hep-ph/0208258})

\bibitem{Shigetani:1993dx}
Shigetani T, Suzuki K and Toki H 1993 {\em Phys. Lett.\/} {\bf B308} 383--388
  (\textit{Preprint} \eprint{hep-ph/9402286})

\bibitem{Davidson:1995uv}
Davidson R~M and Ruiz~Arriola E 1995 {\em Phys. Lett.\/} {\bf B348} 163--169

\bibitem{Weigel:1999pc}
Weigel H, Ruiz~Arriola E and Gamberg L~P 1999 {\em Nucl. Phys.\/} {\bf B560}
  383--427 (\textit{Preprint} \eprint{hep-ph/9905329})

\bibitem{Bentz:1999gx}
Bentz W, Hama T, Matsuki T and Yazaki K 1999 {\em Nucl. Phys.\/} {\bf A651}
  143--173 (\textit{Preprint} \eprint{hep-ph/9901377})

\bibitem{Wijesooriya:2005ir}
Wijesooriya K, Reimer P~E and Holt R~J 2005 {\em Phys. Rev.\/} {\bf C72} 065203
  (\textit{Preprint} \eprint{nucl-ex/0509012})

\bibitem{PhysRevD.16.2219}
Collins J~C and Soper D~E 1977 {\em Phys. Rev. D\/} {\bf 16}(7) 2219--2225

\bibitem{Gottfried:1964nx}
Gottfried K and Jackson J~D 1964 {\em Nuovo Cim.\/} {\bf 33} 309--330

\bibitem{PhysRevD.21.2712}
Lam C~S and Tung W~K 1980 {\em Phys. Rev. D\/} {\bf 21}(9) 2712--2715

\bibitem{PhysRevLett.22.156}
Callan C~G and Gross D~J 1969 {\em Phys. Rev. Lett.\/} {\bf 22}(4) 156--159

\bibitem{Boer:2006eq}
Boer D and Vogelsang W 2006 {\em Phys. Rev.\/} {\bf D74} 014004
  (\textit{Preprint} \eprint{hep-ph/0604177})

\bibitem{Zhu:2006gx}
Zhu L~Y {\em et~al.\/} (FNAL-E866/NuSea) 2006  (\textit{Preprint}
  \eprint{hep-ex/0609005})

\bibitem{Brandenburg:1993cj}
Brandenburg A, Nachtmann O and Mirkes E 1993 {\em Z. Phys.\/} {\bf C60}
  697--710

\bibitem{PhysRevD.60.014012}
Boer D 1999 {\em Phys. Rev. D\/} {\bf 60}(1) 014012

\bibitem{PhysRevD.57.5780}
Boer D and Mulders P~J 1998 {\em Phys. Rev. D\/} {\bf 57}(9) 5780--5786

\bibitem{Collins:1992kk}
Collins J~C 1993 {\em Nucl. Phys.\/} {\bf B396} 161--182 (\textit{Preprint}
  \eprint{hep-ph/9208213})

\bibitem{Lu:2005rq}
Lu Z and Ma B~Q 2005 {\em Phys. Lett.\/} {\bf B615} 200--206 (\textit{Preprint}
  \eprint{hep-ph/0504184})

\bibitem{PhysRevLett.39.1116}
Close F~E and Sivers D 1977 {\em Phys. Rev. Lett.\/} {\bf 39}(18) 1116--1120

\bibitem{Ralston:1979ys}
Ralston J~P and Soper D~E 1979 {\em Nucl. Phys.\/} {\bf B152} 109

\bibitem{fair}
Gutbrod H~H, Augustin I, Eickhoff H, Gross K~D, Henning W~F, Krämer D and
  Walter G, eds 2006 {\em {FAIR} Baseline Technical Report\/}

\bibitem{RHICII}
 2006 Future science at the relativistic heavy ion collider Tech. Rep.
  BNL-77334-2006-IR Brookhaven National Laboratory
  http://www.bnl.gov/physics/rhicIIscience

\bibitem{Perdekamp}
Perdekamp M~G 2006 in {\em 2nd Workshop on the QCD Structure of the Nucleon
  (QCDN06)\/} http://www.lnf.infn.it/conference/qcdn06

\bibitem{Mulders:1995dh}
Mulders P~J and Tangerman R~D 1996 {\em Nucl. Phys.\/} {\bf B461} 197--237
  (\textit{Preprint} \eprint{hep-ph/9510301})

\bibitem{PhysRevLett.67.552}
Jaffe R~L and Ji X 1991 {\em Phys. Rev. Lett.\/} {\bf 67}(5) 552--555

\bibitem{Barone:2003fy}
Barone V and Ratcliffe P~G 2003 {\em Transverse spin physics\/} (World
  Scientific)

\bibitem{Barone:2001sp}
Barone V, Drago A and Ratcliffe P~G 2002 {\em Phys. Rept.\/} {\bf 359} 1--168
  (\textit{Preprint} \eprint{hep-ph/0104283})

\bibitem{Jaffe:1991ra}
Jaffe R~L and Ji X~D 1992 {\em Nucl. Phys.\/} {\bf B375} 527--560

\bibitem{PhysRevD.41.83}
Sivers D 1990 {\em Phys. Rev. D\/} {\bf 41}(1) 83--90

\bibitem{PhysRevD.43.261}
Sivers D 1991 {\em Phys. Rev. D\/} {\bf 43}(1) 261--263

\bibitem{PhysRevLett.77.2626}
Bravar A, Adams D~L, Akchurin N, Belikov N~I, Bonner B~E, Bystricky J, Corcoran
  M~D, Cossairt J~D, Cranshaw J, Derevschikov A~A, En'yo H, Funahashi H, Goto
  Y, Grachov O~A, Grosnick D~P, Hill D~A, Iijima T, Imai K, Itow Y, Iwatani K,
  Kharlov Y~V, Kuroda K, Laghai M, Lehar F, de~Lesquen A, Lopiano D and
  Luehring F~C 1996 {\em Phys. Rev. Lett.\/} {\bf 77}(13) 2626--2629

\bibitem{PhysRevLett.41.1689}
Kane G~L, Pumplin J and Repko W 1978 {\em Phys. Rev. Lett.\/} {\bf 41}(25)
  1689--1692

\bibitem{Collins:2002kn}
Collins J~C 2002 {\em Phys. Lett.\/} {\bf B536} 43--48 (\textit{Preprint}
  \eprint{hep-ph/0204004})

\bibitem{Brodsky:2002cx}
Brodsky S~J, Hwang D~S and Schmidt I 2002 {\em Phys. Lett.\/} {\bf B530}
  99--107 (\textit{Preprint} \eprint{hep-ph/0201296})

\bibitem{Efremov:1984ip}
Efremov A~V and Teryaev O~V 1985 {\em Phys. Lett.\/} {\bf B150} 383

\bibitem{PhysRevLett.67.2264}
Qiu J and Sterman G 1991 {\em Phys. Rev. Lett.\/} {\bf 67}(17) 2264--2267

\bibitem{Ji:2006br}
Ji X, Qiu J~W, Vogelsang W and Yuan F 2006 {\em Phys. Lett.\/} {\bf B638}
  178--186 (\textit{Preprint} \eprint{hep-ph/0604128})

\bibitem{Ji:2006vf}
Ji X, Qiu J~W, Vogelsang W and Yuan F 2006 {\em Phys. Rev.\/} {\bf D73} 094017
  (\textit{Preprint} \eprint{hep-ph/0604023})

\bibitem{Airapetian:2004tw}
Airapetian A {\em et~al.\/} (HERMES) 2005 {\em Phys. Rev. Lett.\/} {\bf 94}
  012002 (\textit{Preprint} \eprint{hep-ex/0408013})

\bibitem{Anselmino:2005ea}
Anselmino M {\em et~al.\/} 2005 {\em Phys. Rev.\/} {\bf D72} 094007
  (\textit{Preprint} \eprint{hep-ph/0507181})

\bibitem{Vogelsang:2005cs}
Vogelsang W and Yuan F 2005 {\em Phys. Rev.\/} {\bf D72} 054028
  (\textit{Preprint} \eprint{hep-ph/0507266})

\bibitem{Alexakhin:2005iw}
Alexakhin V~Y {\em et~al.\/} (COMPASS) 2005 {\em Phys. Rev. Lett.\/} {\bf 94}
  202002 (\textit{Preprint} \eprint{hep-ex/0503002})

\bibitem{Collins:2005rq}
Collins J~C {\em et~al.\/} 2006 {\em Phys. Rev.\/} {\bf D73} 094023
  (\textit{Preprint} \eprint{hep-ph/0511272})

\bibitem{Ashman:1987hv}
Ashman J {\em et~al.\/} (European Muon) 1988 {\em Phys. Lett.\/} {\bf B206} 364

\bibitem{Ashman:1989ig}
Ashman J {\em et~al.\/} (European Muon) 1989 {\em Nucl. Phys.\/} {\bf B328} 1

\bibitem{PhysRevD.58.112002}
Adeva B, Akdogan T, Arik E, Badelek B, Bardin G, Baum G, Berglund P, Betev L,
  Birsa R, de~Botton N, Bradamante F, Bravar A, Bressan A, B\"ultmann S, Burtin
  E, Cavata C, Crabb D, Cranshaw J, \ifmmode~\mbox{\c{C}}\else \c{C}\fi{}uhadai
  T, Dalla~Torre S, van Dantzig R, Derro B, Deshpande A, Dhawan S, Dulya C,
  Eichblatt S and Fasching D 1998 {\em Phys. Rev. D\/} {\bf 58}(11) 112002

\bibitem{Airapetian:2004mi}
Airapetian A {\em et~al.\/} (HERMES) 2005 {\em Phys. Rev.\/} {\bf D71} 032004
  (\textit{Preprint} \eprint{hep-ex/0412027})

\bibitem{Dressler:2000xj}
Dressler B, Goeke K, Schweitzer P, Weiss C and Polyakov M~V 2000 {\em Prog.
  Part. Nucl. Phys.\/} {\bf 44} 293--303

\bibitem{Dressler:1999zg}
Dressler B, Goeke K, Polyakov M~V and Weiss C 2000 {\em Eur. Phys. J.\/} {\bf
  C14} 147--157 (\textit{Preprint} \eprint{hep-ph/9909541})

\bibitem{Dressler:1999zv}
Dressler B {\em et~al.\/} 2001 {\em Eur. Phys. J.\/} {\bf C18} 719--722
  (\textit{Preprint} \eprint{hep-ph/9910464})

\end{thebibliography}

\end{document}